\documentclass[aps,prl,amsmath,reprint,amssymb,superscriptaddress]{revtex4-2}
\usepackage{graphicx}

\begin{document}
\preprint{Version v1}% insert preprint number here
%
%%%%%%%%%%%%%%%%%%%%%%%%%%%%%%%%%%%%%%%%%%%%%%%%%%%%%%%%%%%%%%%%%%%%
\title{A robust weak topological insulator in a bismuth halide Bi$_4$Br$_2$I$_2$}
\author{Ryo~Noguchi}
\altaffiliation[Present address: ]{Department of Physics and Institute for Quantum Information and Matter, California Institute of Technology, Pasadena, California 91125, USA}
\affiliation{Institute for Solid State Physics (ISSP), University of Tokyo, Kashiwa, Chiba 277-8581, Japan}

\author{Masaru~Kobayashi} 
\affiliation{Materials and Structures Laboratory, Tokyo Institute of Technology, Yokohama, Kanagawa 226-8503, Japan}

\author{Kaishu~Kawaguchi}
\affiliation{Institute for Solid State Physics (ISSP), University of Tokyo, Kashiwa, Chiba 277-8581, Japan}

\author{Chun~Lin}
\affiliation{Institute for Solid State Physics (ISSP), University of Tokyo, Kashiwa, Chiba 277-8581, Japan}

\author{Hiroaki~Tanaka}
\affiliation{Institute for Solid State Physics (ISSP), University of Tokyo, Kashiwa, Chiba 277-8581, Japan}

\author{Kenta~Kuroda}
\altaffiliation[Present address: ]{Graduate School of Advanced Science and Engineering, Hiroshima University, 1-3-1 Kagamiyama, Higashi-hiroshima 739-8526, Japan}
\affiliation{Institute for Solid State Physics (ISSP), University of Tokyo, Kashiwa, Chiba 277-8581, Japan}

\author{Ayumi~Harasawa}
\affiliation{Institute for Solid State Physics (ISSP), University of Tokyo, Kashiwa, Chiba 277-8581, Japan}

\author{Viktor~Kandyba}
\affiliation{Elettra - Sincrotrone Trieste, S.S.14, 163.5 km, Basovizza, Trieste, Italy}

\author{Mattia~Cattelan}
\affiliation{Elettra - Sincrotrone Trieste, S.S.14, 163.5 km, Basovizza, Trieste, Italy}

\author{Alexei~Barinov}
\affiliation{Elettra - Sincrotrone Trieste, S.S.14, 163.5 km, Basovizza, Trieste, Italy}

\author{Makoto~Hashimoto}
\affiliation{Stanford Synchrotron Radiation Light source, SLAC National Accelerator Laboratory, 2575 Sand Hill Road, Menlo Park, California 94025, USA}

\author{Donghui~Lu}
\affiliation{Stanford Synchrotron Radiation Light source, SLAC National Accelerator Laboratory, 2575 Sand Hill Road, Menlo Park, California 94025, USA}

\author{Takao~Sasagawa} 
\email{sasagawa.t.aa@m.titech.ac.jp}
\affiliation{Materials and Structures Laboratory, Tokyo Institute of Technology, Yokohama, Kanagawa 226-8503, Japan}

\author{Takeshi~Kondo}
\email{kondo1215@issp.u-tokyo.ac.jp}
\affiliation{Institute for Solid State Physics (ISSP), University of Tokyo, Kashiwa, Chiba 277-8581, Japan}
\affiliation{Trans-scale Quantum Science Institute, The University of Tokyo, Tokyo 113-0033, Japan}

%%%%%%%%%%%%%%%%%%%%%%%%%%%%%%%%%%%%%%%%%%%%%%%%%%%%%%%%%%%%%%%%%%%%
\begin{abstract}                                                   %
%%%%%%%%%%%%%%%%%%%%%%%%%%%%%%%%%%%%%%%%%%%%%%%%%%%%%%%%%%%%%%%%%%%%
%
We apply a topological material design concept for selecting a bulk topology of 3D crystals by different van-der-Waals stacking of 2D topological insulator layers, and find a bismuth halide Bi$_4$Br$_2$I$_2$ to be an ideal weak topological insulator (WTI) with the largest band gap ($\sim$230 meV) among all the WTI candidates, by means of angle-resolved photoemission spectroscopy (ARPES), density functional theory (DFT) calculations, and resistivity measurements. Our results vastly expand future opportunities for fundamental research and device applications with a robust WTI.

\end{abstract}
\maketitle
%
% Introduction %

\begin{figure*}[htb]
\begin{center}
\includegraphics[width=1\textwidth]{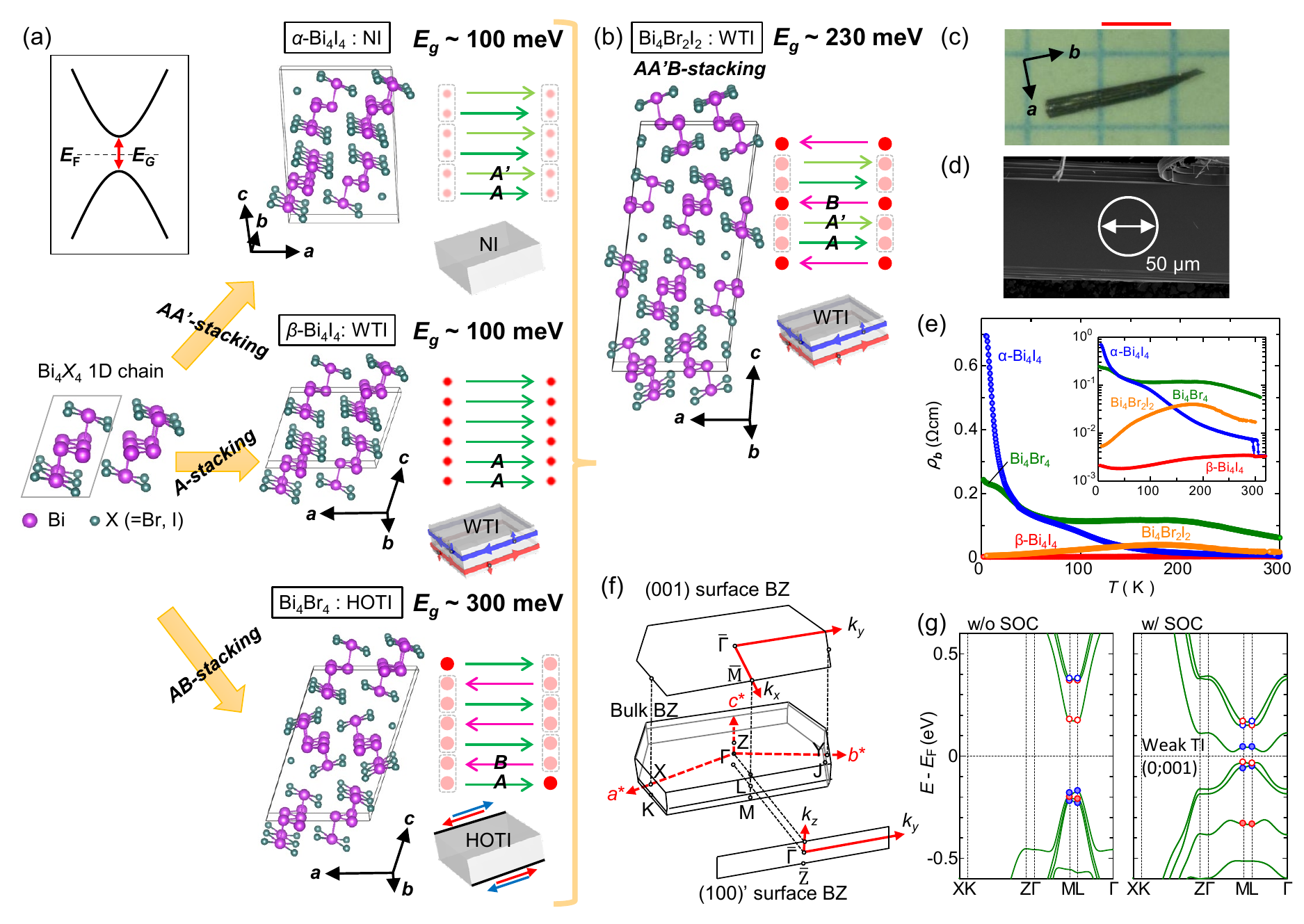}
\caption[]{Topological material design concept to construct a robust WTI with a large band gap $E_g$. (a) Topological phases selected by different van-der-Waals stackings of Bi$_4 X_4$ ($X$=Br, I) chains. A NI, a WTI, and a HOTI have been experimentally validated in $A$-stacking ($\alpha$-Bi$_4$I$_4$ with $E_g$$\sim$100 meV), $AA'$-stacking ($\beta$-Bi$_4$I$_4$ with $E_g$$\sim$100 meV), and $AB$-stacking (Bi$_4$Br$_4$ $E_g$$\sim$300 meV), respectively. (b) The trilayer Bi$_4$Br$_2$I$_2$ built from $AA'B$-stacking, taking advantage both of $\beta$-Bi$_4$I$_4$ with a WTI state and Bi$_4$Br$_4$ with a large band gap. Topological spin currents and the annihilated ones are expressed by densely and lightly painted circles. (c) Photograph of a Bi$_4$Br$_2$I$_2$ crystal. The scale bar is 1 mm. (d) Scanning electron microscope (SEM) image of a cleaved surface. (e) The electrical resistivity of Bi$_4$Br$_2$I$_2$, compared with that for $\alpha$-Bi$_4$I$_4$, $\beta$-Bi$_4$I$_4$, and Bi$_4$Br$_4$. (f) Brillouin zone (BZ) for the bulk and projected surfaces of $AA'B$-stacked Bi$_4 X_4$. (g) Bulk band calculations of $AA'B$-stacked Bi$_4Br_4$ with and without SOC. Here, the halogen atoms $X$ are all set as Br, rather than a mix of half Br and half I, to simplify the calculation. The red and blue circles indicate the even parities and odd parities, respectively. }
\label{fig1}
\end{center}
\end{figure*}

% previous paper

\begin{figure*}[htb]
\begin{center}
\includegraphics[width=1\textwidth]{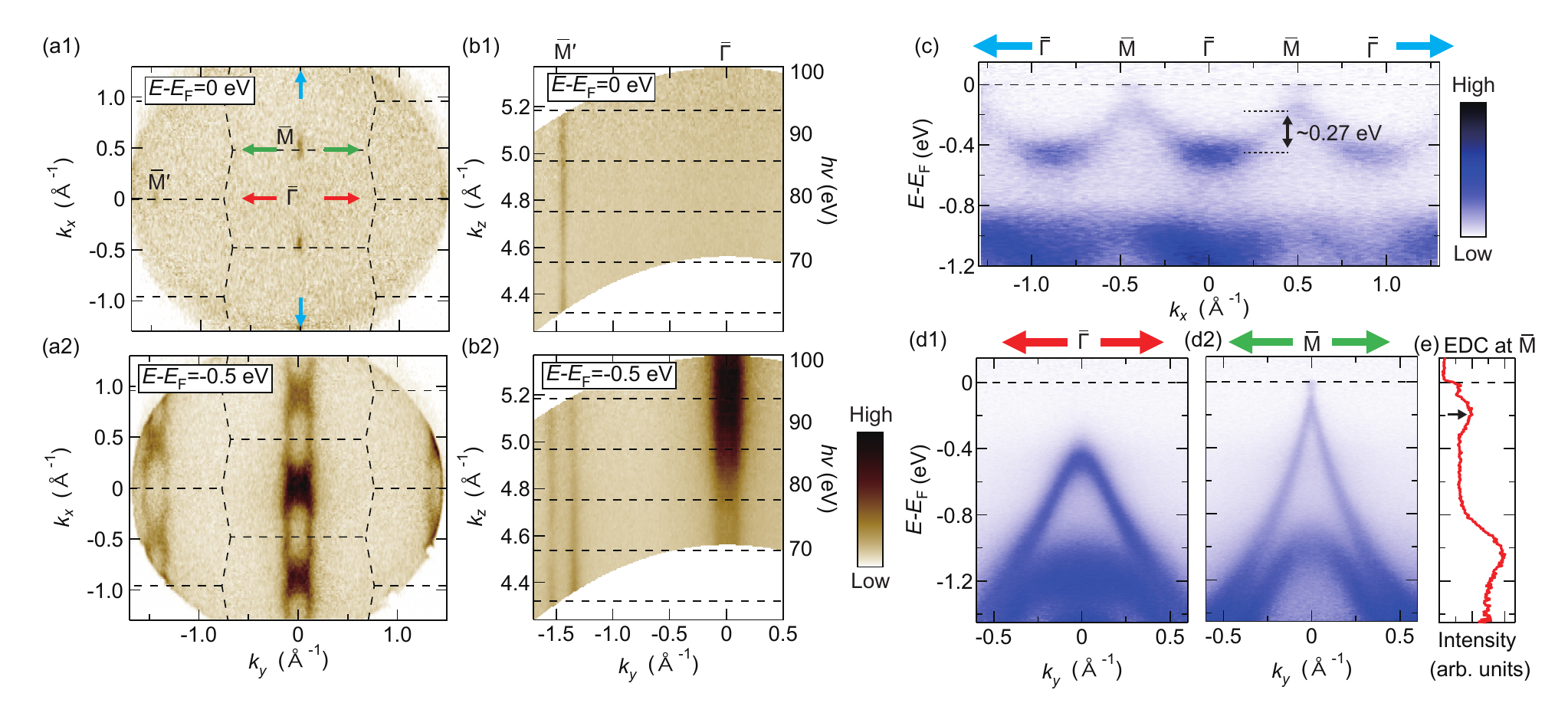}
\caption[]{Synchrotron-ARPES results revealing the bulk band of Bi$_4$Br$_2$I$_2$. 
 (a1,a2) ARPES maps along a $k_y - k_x$ sheet at $E-E_{\rm F}$ = 0 and $-0.5~{\rm eV}$, respectively, measured at $h\nu =100~{\rm eV} $. The black dashed lines indicate the surface BZ for the (001) surface. (b1,b2) ARPES maps for a $k_y -k_z$ sheet at $E-E_{\rm F}$ = 0 and $-0.5~{\rm eV}$, respectively; the $k_z$ dependence is obtained along $\bar{\rm M}-\bar{\Gamma}$ by varying $h\nu$ from 70 to 100 eV. Bulk band dispersions along the $\bar{\rm M}-\bar{\Gamma}-\bar{\rm M}$ cut (c) and across $\bar{\Gamma}$ and  $\bar{\rm M}$ (d1,d2), measured at $h\nu=100~{\rm eV}$. (e) EDC at $\bar{\rm M}$ extracted from (d2). The black arrow indicates the valence band top.
}
\label{fig2}
\end{center}
\end{figure*}
% Experimental

%
%%%%%%%%%%%%%%%%%%%%%%%%%%%%%%%%%%%%%%%%%%%%%%%%%%%%%%%%%%%%%%%%%%%%
% Results                                                          %
%%%%%%%%%%%%%%%%%%%%%%%%%%%%%%%%%%%%%%%%%%%%%%%%%%%%%%%%%%%%%%%%%%%%
%
\begin{figure}[bth]
\begin{center}
\includegraphics[width=1\columnwidth]{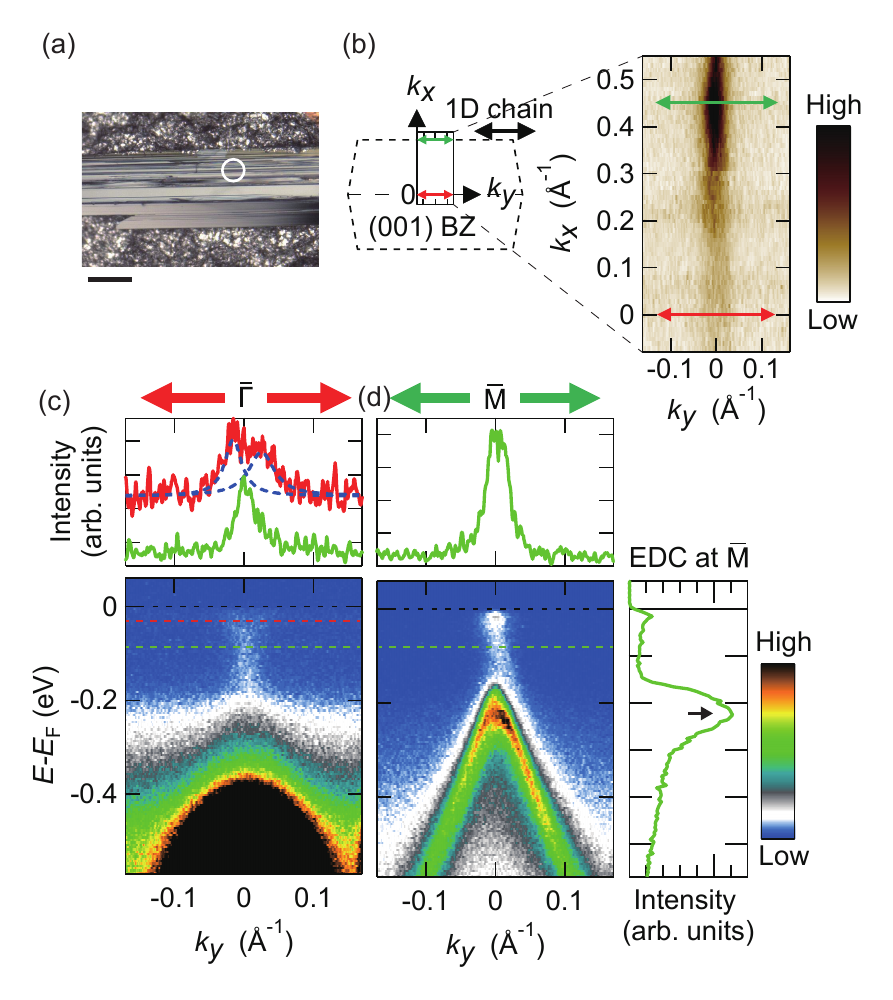}
\caption[]{Laser-ARPES data showing overlapped intensities of the bulk band and the quasi-1D Dirac band on the surface.  (a) Photograph of a cleavage surface measured. The scale bar is 100 $\mu$m. The white circle denotes the position observed in our measurements. (b) ARPES map along the $k_y-k_x$ plane at $E_{\rm F}$. The left shows the (001) surface BZ. (c) ARPES band map across $\bar{\Gamma}$ (bottom) and the corresponding MDCs (top) for two different binding energies (dashed line in the map). Two peak structures in the MDC are fit well by doubled Lorentzian curves (blue dotted lines). (d) ARPES band map across $\bar{\rm M}$ (bottom), the corresponding MDC (top), and EDC at $\bar{\rm M}$ (right). The high-quality spectrum allows defining the bulk band gap as $\sim230$ meV (pointed by the arrow).}
\label{fig3}
\end{center}
\end{figure}

\begin{figure*}[th]
\begin{center}
\includegraphics[width=0.98\textwidth]{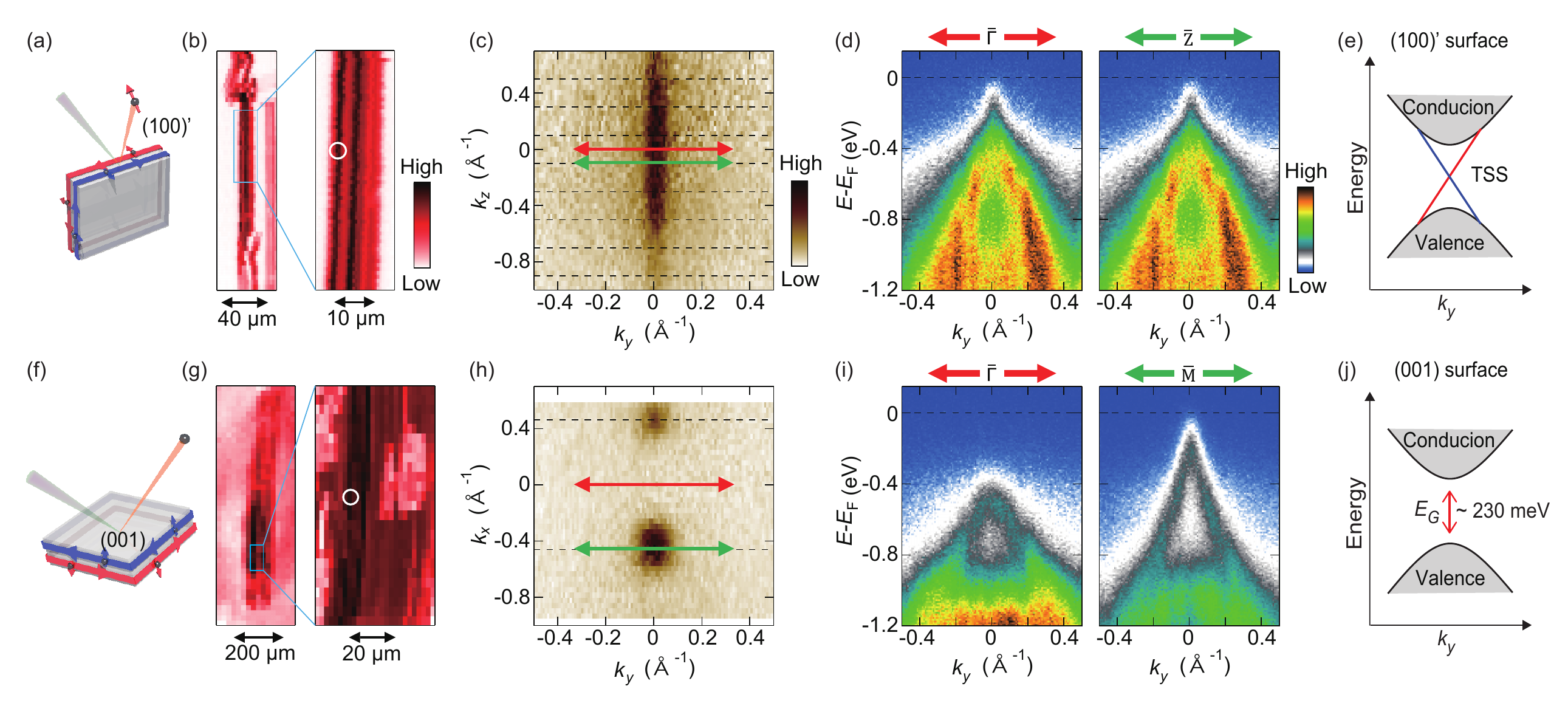}
\caption[]{Selective observation of the top and side surfaces by nano-ARPES.
The upper panels demonstrate the results for the $(100)'$ side surface.
(a) Experimental geometry.
(b) Photoemission intensity map in real space. 
The position measured by nano-ARPES is indicated by a white circle. 
(c) ARPES intensity map at $E_{\rm F}$. 
(d) ARPES band maps across the two momentum cuts depicted in (c). 
(e) Schematic of the bulk and surface bands with TSS (red and blue lines).
 (f-j) The same data as (a-e), but for the $(001)$ top surface. Differently from the side surface, TSS is absent, showing a large band gap [arrow in (j)].
}
\label{fig4}
\end{center}
\end{figure*}

A weak topological insulator (WTI)  flows highly directional spin currents along side surfaces of crystals \cite{Fu2007a}. The topological spin currents would be more robust against impurity scattering than those in a strong topological insulator (STI) that prohibits only perfect backscattering.

WTIs, thus, could be even more advantageous than STIs for various applications. In contrast to STIs, however, the materials hosting WTI states are very scarce \cite{Yan2012, Rasche2013,Tang2014,Yang2014,Liu2016bb,Zhang2017d,Liu2018h,Noguchi2019,Huang2021,Lee2021}. In addition, the WTIs established so far have a band gap relatively smaller than those of the prototype STIs reaching up to $\sim$ 350 meV \cite{Ando2013}. It may have the WTI states fragile against excitations and difficult to extract without being masked by the non-trivial bulk contributions. Therefore, the discovery of a WTI with a large band gap has been awaited in materials science.

A difficulty in searching for WTIs comes from their property that the topological surface state (TSS) resides only on the side surface of crystals that are usually not cleavable, preventing the verification of their bulk band topology.
Another difficulty is finding an insulator, not a semimetal, hosting a WTI state. This is a required condition to utilize TSSs for research and application without contamination of trivial bulk conductivity. So far, only four compounds have been proposed as WTIs via experiments: Bi$_{14}$Rh$_3$I$_9$, $\beta$-Bi$_4$I$_4$, ZrTe$_5$, and HfTe$_5$ \cite{Rasche2013,Noguchi2019,Zhang2017d,Liu2018h}. Bi$_{14}$Rh$_3$I$_9$ has a relatively large band gap ($\sim200$ meV), whereas the other three have small ones ($\sim100$, $\sim30$, and $\sim50$ meV each). Bi$_{14}$Rh$_3$I$_9$ is a promising candidate for a WTI with a large gap. However, the side surface of this compound is not cleavable, preventing the direct observation of the topological surface band required for the identification of a WTI.

Recently, quasi-one-dimensional (quasi-1D) bismuth halides Bi$_4 X_4$ ($X$=Br, I) have attracted much attention as a versatile platform to realize various topological phases including WTI \cite{Zhou2014b,Zhou2015a,Liu2016bb,Autes2015b,InsulatorPisoni,Qi2017,Wang2018,Chen2018f,
Deng2019a,Noguchi2019,Li2019d,Hsu2019b,Qiao2021,Noguchi2021a,Wang2021,Zhuang2021,Huang2021,
Peng2021,Chen2022,Shumiya2022,Liu2022,Yang2022,Zhong2022,Zhang2022}. These compounds can be regarded as the stacking of two-dimensional TIs (2D TIs) \cite{Zhou2014b}, and several types of stacking structures have been proposed \cite{VonSchnering1978,Popovkin2001}. 
The bulk topology sensitively depends on the stacking sequences.
Most importantly, the crystals built from chains can be easily cleaved by scotch tape along both top and side surfaces, allowing one to observe topological boundary states residing on the crystal surface or hinge. 

So far, three types of crystal structures ($\alpha$-Bi$_4$I$_4$, $\beta$-Bi$_4$I$_4$, and Bi$_4$Br$_4$) have been investigated by theory and experiments [Fig. 1(a)], and each has been experimentally identified as a normal insulator (NI), a WTI, and a higher-order topological insulator (HOTI). The band gap of the WTI ($\beta$-Bi$_4$I$_4$) is, however, rather small ($\sim$100 meV \cite{Noguchi2019}), compared to that of Bi$_{14}$Rh$_3$I$_9$ ($\sim$200 meV).

In this letter, we find the trilayer Bi$_4$Br$_2$I$_2$ to be a weak topological insulator with a band gap of as large as $\sim$230 meV. The magnitude of the band gap is comparable to or even larger than that ($\sim$200 meV) in Bi$_{14}$Rh$_3$I$_9$. Differently from Bi$_{14}$Rh$_3$I$_9$, Bi$_4$Br$_2$I$_2$ built from chains has the top and side planes both naturally cleavable, fulfilling an indispensable condition to identify a WTI. 
We successfully observe the topological side surface separately from the topologically dark top plane, owing to nano-ARPES which allows their independent observation.

%%%%%%%%%%%%%Method
Single crystals of Bi$_4$Br$_2$I$_2$ were grown by the chemical vapor transport method. Synchrotron-ARPES, laser-ARPES, and nano-ARPES measurements were performed at the beamline 5-2 of Stanford Synchrotron Radiation Lightsource (SSRL), the Institute for Solid State Physics of the University of Tokyo~\cite{Shimojima2015,Yaji2016}, and the 3.2L-Spectromicroscopy beamline of the Elettra Light Source~\cite{Dudin2010}, respectively.
Relativistic first-principles calculations were performed using the \textsc{WIEN2k} code~\cite{Blaha2019} under the conditions (GGA-PBE exchange-correlation functional, mBJ exchange potential, etc.) described elsewhere \cite{Noguchi2019,Noguchi2021a}. The details of the crystal growth, structure analysis, and ARPES setting are explained in the Supplemental Material \cite{supple}.
%%%%%%%%%%%%%
 
We start by explaining the topological material design concept. In $A$-stacking, the $Z_2$ topology of the 2D layers is preserved in the 3D crystal, and the spin current is accumulated to generate a massive flow along the side surface. 
In $AA'$-stacking with double layers slightly shifted from each other, the edge states of paired layers interact, vanishing the spin currents, and a normal insulator state is obtained ($Z_2=0$). 
The $AB$-stacking also consists of double layers. However, one of the double layers is flipped by 180 degrees, and the inversion center comes on the layers, not between them as in the $ AA'$-stacking. 
Consequently, although $Z_2=0$, the band topology is categorized as non-trivial ($Z_4=2$) under the topological classification extended to the $Z_4$ index, which yields a higher-order topological insulator (HOTI) \cite{Tang2019a,Vergniory,Zhang2019l,Hsu2019b,Noguchi2021a}. 
It annihilates the topological spin currents from surfaces but leaves them on the crystal hinge. 
According to these concepts, a NI, a WTI, and a HOTI have been experimentally validated in $\alpha$-Bi$_4$I$_4$ ($A$-stacking), $\beta$-Bi$_4$I$_4$ ($AA'$-stacking), and Bi$_4$Br$_4$ ($AB$-stacking), respectively. 

We propose here that the suitable choice of the component 2D layers is critical for realizing a WTI with a large band gap. Theoretically, the monolayer Bi$_4$Br$_4$ has a larger band gap than the monolayer Bi$_4$I$_4$ \cite{Zhou2014b}. The band gap of the layer-stacked 3D crystal could also be enlarged by substituting I for Br in Bi$_4$I$_4$. 
The Bi$_4$Br$_4$ crystal, indeed, shows a large band gap of $\sim$300 meV \cite{Noguchi2021a}. It is, however, double-layered with the $AB$-stacking, which hosts a HOTI, not a WTI. 
 Bi$_4$Br$_4$ built from a single-layered $A$-stacking has been theoretically suggested as a WTI \cite{Liu2016bb}. Nevertheless, such crystals are not stable and cannot be synthesized. It is, thus, desired to consider an alternative structure realizing a WTI that can be synthesized and still take advantage of Bi$_4$Br$_4$ layers with a large band gap. 

We find $AA'B$-stacking to fulfill such a condition and demonstrate that Bi$_4$Br$_2$I$_2$ with this stacking \cite{supple,Popovkin2001} realizes an ideal WTI with a large gap. This structure can be viewed as an alternative stacking of $AA'$-layers and $B$-layer mutually flipped by 180 degrees. Since the $AA'$-layers should behave as normal insulators, as in $\alpha$-Bi$_4$I$_4$, the $AA'B$-stacked structure can be viewed as a 2D TI layers ($B$-layers) alternatively stacked with insulator blocks ($AA'$-layers). In the band structure, only a single parity inversion gets effective when all the trilayer-split bands are inverted by strong spin-orbit interaction, as calculated in Fig. 1(f). This leads to a similar situation as in the single-layered $\beta$-Bi$_4$I$_4$, where a single band inversion generates a WTI state. 

The resistivities along chains provide two implications [Fig. 1(e)]. (1) The resistivity of Bi$_4$Br$_2$I$_2$ shows an insulating behavior around room temperature. The magnitude comes to between those of Bi$_4$Br$_4$ and Bi$_4$I$_4$, indicating that the bulk gap of Bi$_4$Br$_2$I$_2$ is smaller than that of Bi$_4$Br$_4$ (a HOTI) but larger than $\alpha$- and $\beta$-Bi$_4$I$_4$ (a normal insulator and a WTI, respectively).
(2) The resistivity of Bi$_4$Br$_2$I$_2$ decreases upon cooling below $\sim$150 K and turns metallic at 
low temperatures. The metallic resistivity is much smaller than the low-temperature resistivity of Bi$_4$Br$_4$ (a HOTI) and comparable to (smaller but only slightly than) that of $\beta$-Bi$_4$I$_4$. These results 
imply that Bi$_4$Br$_2$I$_2$ is in a topological phase with a large insulating band gap in bulk and massive currents on the topological surface.
  
%Fig. 2
We performed synchrotron-based ARPES measurements on the surface cleaved along the $ab$ planes by tape.
Typically, the surface composes relatively large terraces larger than the synchrotron light spot ($\sim$50 $\mu$m), together with a bunch of small steps. Figures 2(a1) and 2(a2) plot ARPES intensity maps along a $k_y -k_x$ sheet at different binding energies measured at $h\nu=100$ eV. At the Fermi level ($E_{\rm F}$) [Fig. 2(a1)], island-like weak intensities are repeatedly observed at $\bar{\rm M}$ in the (001) surface Brillouin zones.  Anisotropic features with extended parallel segments are clarified at higher binding energies [$E-E_{\rm F} = -0.5$ eV in Fig. 2(a2)], indicating weak interchain hopping. The quasi-1D feature is further confirmed by investigating the band dispersion in the $k_z$ direction (or the $AA'B$-stacking direction) with changing photon energy.  Figures. 2(b1) and 2(b2) plot ARPES intensities along the $\bar{\rm M}$-$\bar{\rm \Gamma}$ cut [red arrow in Fig. 2(a1)] at photon energies from 70 to 100 eV  [$E-E_{\rm F} = 0$ eV and $-0.5$ eV in Figs. 2(a1) and 2(a2), respectively]. Although there is some intensity modulation due to the matrix element effects, we find almost no dispersion along $k_z$ in agreement with calculations [Fig. 1(g)].

In Fig. 2(c) and 2(d1,d2), we examine the ARPES energy dispersions.
The bulk valence band disperses in energy by $\sim$270 meV along the $\bar{\rm M}-\bar{\Gamma}-\bar{\rm M}$ direction [Fig. 2(c)]. This is similar to the corresponding dispersion of Bi$_4$I$_4$ ($\sim$260 meV) and smaller than that of Bi$_4$Br$_4$ ($\sim$350 meV) \cite{Noguchi2019,Noguchi2021a}. The energy distribution curve (EDC) at $\bar{\rm M}$ [Fig. 2(e)] indicates that the top of the bulk valence band is deep below $E_{\rm F}$ (marked by an arrow), whereas the bulk conduction band is located above $E_{\rm F}$ and exhibits a spectral tail with weak intensities which forms the islands in the Fermi surface map [Fig. 2(a1)]. 

%Fig. 3
We also use laser-based ARPES with high energy and momentum resolutions. This technique has enabled the direct observation of in-gap states in various topological materials \cite{Lv2019a}. The cleaved surface is confirmed to expose both the (001) top and (100) side planes [Fig. 3(a)]. 
To detect the signals of the side surface where TSSs are expected to exist, 
we illuminate a laser with a 50 $\mu$m spot onto a surface portion with many steps marked by a circle in Fig. 3(a).
Figure 3(b) plots the obtained Fermi surface map for the rectangular region depicted in the left panel.
Importantly, we observe a quasi-1D structure extending over the entire momentum space measured, although it is overlapped by island-like strong intensities around $\bar{\rm M}$ coming from the spectral tail of the bulk conduction band above $E_F$. We find this quasi-1D band forms a Dirac-like dispersion, as demonstrated in Fig. 3(c,d). Two peaks are observed in the momentum distribution curve (MDC) near $E_{\rm F}$ [a red curve in the top panel of Fig. 3(c)], away from $\bar{\rm M}$ dominated by the bulk conduction band in intensity. 
The two peaks merge into one peak at $E-E_{\rm F} \sim 100$ meV, expected to be a Dirac point  [green curves in the top panels of Figs. 3(c,d)]. The quasi-1D Dirac band is observed inside the bulk band gap, so it is attributed to be TSSs. 

The excellent energy resolution in laser-ARPES allows one to estimate the gap magnitude as large as $\sim$230 meV from the EDC at $\bar{\rm M}$ [right panel of Fig. 3(d)]. We should note, however, that this value is only the lower limit of the gap size since the valence band is situated above $E_F$.
This gap is smaller than that of Bi$_4$Br$_4$ ($\sim$300 meV), but it should be more than twice larger than that of Bi$_4$I$_4$ ($\sim$100 meV). Bi$_4$Br$_2$I$_2$ containing Br and I, therefore, realizes a stable and robust WTI state by taking advantage of both $\beta$-Bi$_4$I$_4$ (a WTI) and Bi$_4$Br$_4$ (a large band gap).
 
%Fig. 4
To confirm the side surface origin for the quasi-1D in-gap states, we employed nano-ARPES with a focused photon beam less than 1 $\mu$m spot size, which allows selectively observing the (100) side and (001) top planes. The experimental geometries are drawn in Fig. 4(a) and Fig. 4(f). The sample crystals have the typical size of 0.2 mm$\times$2 mm$\times$20 $\mu$m each in the $a$-, $b$-, and $c$-axis direction.  We cleaved those along the $bc$- and $ab$-surfaces by tape to dominantly expose the (001) and (100) planes and independently observed each plane by nano-ARPES. 

Figure 4(b) shows the microscopic real-space intensity map of the Bi $5d$ core level obtained for the cleavage surface along the side $bc$-plane. Several needle-like strong intensities are seen along chains, confirming that the very narrow (100) plane is successfully cleaved. We select the brightest spot [a white circle in Fig. 4(b)] for the nano-ARPES measurements. The obtained Fermi surface map [Fig. 4(c)] shows a quasi-1D feature along $k_z$ over many Brillouin zones. The ARPES energy dispersions [Fig. 4(d)] exhibit metallic in-gap states of quasi-1D structures with almost no variation between the two momentum cuts across the zone center and corner [red and green lines in Fig. 4(c), respectively]. 

In contrast, we obtained high intensities over a large area on the top $ab$-plane [Fig. 4(f,g)]. 
Similarly to synchrotron-based data (Fig. 2), island-like intensities are observed in the ARPES map at $E_{\rm F}$ with a periodicity for the (001) surface Brillouin zone [Fig. 4(h)]. The ARPES dispersions [Fig. 4(i)] capture the bulk valence band with a clear difference between the zone center and corner cuts [red and green lines in Fig. 4(h)], and the quasi-1D states observed on the side surface are absent. These nano-ARPES data are consistent with the theoretical prediction that Bi$_4$Br$_2$I$_2$ is a WTI. That is, the top surface exhibits a gapped insulating band without TSS, whereas the side surface hosts a gapless metallic quasi-1D TSS, as summarized in Figs. 4(j) and 4(e), respectively.

%%%%%%%%%%%%%%%%%%%%%%%%%%%%%%%%%%%%%%%%%%%%%%%%%%%%%%%%%%%%%%%%%%%%
% Conclusion %
%%%%%%%%%%%%%%%%%%%%%%%%%%%%%%%%%%%%%%%%%%%%%%%%%%%%%%%%%%%%%%%%%%%%
%
In conclusion, we investigated the electronic structure of a trilayer bismuth halide Bi$_4$Br$_2$I$_2$, based on the topological material design that a half Br and a half I in the anion site bring both advantages of a WTI state in $\beta$-Bi$_4$I$_4$ and a large band gap of Bi$_4$Br$_4$. By selectively observing the top and side surfaces, Bi$_4$Br$_2$I$_2$ was identified as the most robust WTI with the largest band gap ($\sim$230 meV) among the existing WTIs, or at least a similar gap value as in Bi$_{14}$Rh$_3$I$_9$ ($\sim$200 meV). Very importantly, Bi$_4$Br$_2$I$_2$ has advantages overcoming Bi$_{14}$Rh$_3$I$_9$ mainly in two ways due to its unique crystal structure built from chains with the top and side planes both naturally cleavable. One is that the topological state was identified via direct observation of the topological surface band. Second is that the exfoliation technique having got popular in modern research can be employed, leading to various applications. 
This compound, therefore, provides a new excellent platform for future research and device applications utilizing highly directional, massive spin currents.

%
%%%%%%%%%%%%%%%%%%%%%%%%%%%%%%%%%%%%%%%%%%%%%%%%%%%%%%%%%%%%%%%%%%%%
% Acknowledgements                                                 %
%%%%%%%%%%%%%%%%%%%%%%%%%%%%%%%%%%%%%%%%%%%%%%%%%%%%%%%%%%%%%%%%%%%%
%
This work was supported by the JSPS KAKENHI (Grant Nos. JP21H04652, JP21K18181, JP21H04439, JP21H04439, and JP21H05236), by MEXT Q-LEAP (Grant No. JPMXS0118068681), and by MEXT as “Program for Promoting Researches on the Supercomputer Fugaku” (Basic Science for Emergence and Functionality in Quantum Matter Innovative Strongly-Correlated Electron Science by Integration of “Fugaku” and Frontier Experiments, JPMXP1020200104) (Project ID: hp200132/hp210163/hp220166). Use of the Stanford Synchrotron Radiation Lightsource at the SLAC National Accelerator Laboratory is supported by the U.S. Department of Energy, Office of Science, Office of Basic Energy Sciences under contract no. DE-AC02-76SF00515. The nano-ARPES experiments were performed with the approval of Elettra Sincrotrone Trieste (Proposal No. 20210361). R.N. acknowledges support by JSPS under KAKENHI Grant No. JP18J21892 and support by JSPS through the Program for Leading Graduate Schools (ALPS).

%
%%%%%%%%%%%%%%%%%%%%%%%%%%%%%%%%%%%%%%%%%%%%%%%%%%%%%%%%%%%%%%%%%%%%
% References                                                       %
%%%%%%%%%%%%%%%%%%%%%%%%%%%%%%%%%%%%%%%%%%%%%%%%%%%%%%%%%%%%%%%%%%%%

%


\begin{thebibliography}{44}%
\makeatletter
\providecommand \@ifxundefined [1]{%
 \@ifx{#1\undefined}
}%
\providecommand \@ifnum [1]{%
 \ifnum #1\expandafter \@firstoftwo
 \else \expandafter \@secondoftwo
 \fi
}%
\providecommand \@ifx [1]{%
 \ifx #1\expandafter \@firstoftwo
 \else \expandafter \@secondoftwo
 \fi
}%
\providecommand \natexlab [1]{#1}%
\providecommand \enquote  [1]{``#1''}%
\providecommand \bibnamefont  [1]{#1}%
\providecommand \bibfnamefont [1]{#1}%
\providecommand \citenamefont [1]{#1}%
\providecommand \href@noop [0]{\@secondoftwo}%
\providecommand \href [0]{\begingroup \@sanitize@url \@href}%
\providecommand \@href[1]{\@@startlink{#1}\@@href}%
\providecommand \@@href[1]{\endgroup#1\@@endlink}%
\providecommand \@sanitize@url [0]{\catcode `\\12\catcode `\$12\catcode
  `\&12\catcode `\#12\catcode `\^12\catcode `\_12\catcode `\%12\relax}%
\providecommand \@@startlink[1]{}%
\providecommand \@@endlink[0]{}%
\providecommand \url  [0]{\begingroup\@sanitize@url \@url }%
\providecommand \@url [1]{\endgroup\@href {#1}{\urlprefix }}%
\providecommand \urlprefix  [0]{URL }%
\providecommand \Eprint [0]{\href }%
\providecommand \doibase [0]{https://doi.org/}%
\providecommand \selectlanguage [0]{\@gobble}%
\providecommand \bibinfo  [0]{\@secondoftwo}%
\providecommand \bibfield  [0]{\@secondoftwo}%
\providecommand \translation [1]{[#1]}%
\providecommand \BibitemOpen [0]{}%
\providecommand \bibitemStop [0]{}%
\providecommand \bibitemNoStop [0]{.\EOS\space}%
\providecommand \EOS [0]{\spacefactor3000\relax}%
\providecommand \BibitemShut  [1]{\csname bibitem#1\endcsname}%
\let\auto@bib@innerbib\@empty
%</preamble>
\bibitem [{\citenamefont {Fu}\ \emph {et~al.}(2007)\citenamefont {Fu},
  \citenamefont {Kane},\ and\ \citenamefont {Mele}}]{Fu2007a}%
  \BibitemOpen
  \bibfield  {author} {\bibinfo {author} {\bibfnamefont {L.}~\bibnamefont
  {Fu}}, \bibinfo {author} {\bibfnamefont {C.~L.}\ \bibnamefont {Kane}},\ and\
  \bibinfo {author} {\bibfnamefont {E.~J.}\ \bibnamefont {Mele}},\ }\href
  {https://doi.org/10.1103/PhysRevLett.98.106803} {\bibfield  {journal}
  {\bibinfo  {journal} {Phys. Rev. Lett.}\ }\textbf {\bibinfo {volume} {98}},\
  \bibinfo {pages} {106803} (\bibinfo {year} {2007})}\BibitemShut {NoStop}%
\bibitem [{\citenamefont {Yan}\ \emph {et~al.}(2012)\citenamefont {Yan},
  \citenamefont {M{\"{u}}chler},\ and\ \citenamefont {Felser}}]{Yan2012}%
  \BibitemOpen
  \bibfield  {author} {\bibinfo {author} {\bibfnamefont {B.}~\bibnamefont
  {Yan}}, \bibinfo {author} {\bibfnamefont {L.}~\bibnamefont {M{\"{u}}chler}},\
  and\ \bibinfo {author} {\bibfnamefont {C.}~\bibnamefont {Felser}},\ }\href
  {https://doi.org/10.1103/PhysRevLett.109.116406} {\bibfield  {journal}
  {\bibinfo  {journal} {Phys. Rev. Lett.}\ }\textbf {\bibinfo {volume} {109}},\
  \bibinfo {pages} {116406} (\bibinfo {year} {2012})}\BibitemShut {NoStop}%
\bibitem [{\citenamefont {Rasche}\ \emph {et~al.}(2013)\citenamefont {Rasche},
  \citenamefont {Isaeva}, \citenamefont {Ruck}, \citenamefont {Borisenko},
  \citenamefont {Zabolotnyy}, \citenamefont {B{\"{u}}chner}, \citenamefont
  {Koepernik}, \citenamefont {Ortix}, \citenamefont {Richter},\ and\
  \citenamefont {van~den Brink}}]{Rasche2013}%
  \BibitemOpen
  \bibfield  {author} {\bibinfo {author} {\bibfnamefont {B.}~\bibnamefont
  {Rasche}}, \bibinfo {author} {\bibfnamefont {A.}~\bibnamefont {Isaeva}},
  \bibinfo {author} {\bibfnamefont {M.}~\bibnamefont {Ruck}}, \bibinfo {author}
  {\bibfnamefont {S.}~\bibnamefont {Borisenko}}, \bibinfo {author}
  {\bibfnamefont {V.}~\bibnamefont {Zabolotnyy}}, \bibinfo {author}
  {\bibfnamefont {B.}~\bibnamefont {B{\"{u}}chner}}, \bibinfo {author}
  {\bibfnamefont {K.}~\bibnamefont {Koepernik}}, \bibinfo {author}
  {\bibfnamefont {C.}~\bibnamefont {Ortix}}, \bibinfo {author} {\bibfnamefont
  {M.}~\bibnamefont {Richter}},\ and\ \bibinfo {author} {\bibfnamefont
  {J.}~\bibnamefont {van~den Brink}},\ }\href
  {https://doi.org/10.1038/nmat3570} {\bibfield  {journal} {\bibinfo  {journal}
  {Nat. Mater.}\ }\textbf {\bibinfo {volume} {12}},\ \bibinfo {pages} {422}
  (\bibinfo {year} {2013})}\BibitemShut {NoStop}%
\bibitem [{\citenamefont {Tang}\ \emph {et~al.}(2013)\citenamefont {Tang},
  \citenamefont {Yan}, \citenamefont {Cao}, \citenamefont {Wu}, \citenamefont
  {Felser},\ and\ \citenamefont {Duan}}]{Tang2014}%
  \BibitemOpen
  \bibfield  {author} {\bibinfo {author} {\bibfnamefont {P.}~\bibnamefont
  {Tang}}, \bibinfo {author} {\bibfnamefont {B.}~\bibnamefont {Yan}}, \bibinfo
  {author} {\bibfnamefont {W.}~\bibnamefont {Cao}}, \bibinfo {author}
  {\bibfnamefont {S.-C.}\ \bibnamefont {Wu}}, \bibinfo {author} {\bibfnamefont
  {C.}~\bibnamefont {Felser}},\ and\ \bibinfo {author} {\bibfnamefont
  {W.}~\bibnamefont {Duan}},\ }\href
  {https://doi.org/10.1103/PhysRevB.89.041409} {\bibfield  {journal} {\bibinfo
  {journal} {Phys. Rev. B}\ }\textbf {\bibinfo {volume} {89}},\ \bibinfo
  {pages} {041409} (\bibinfo {year} {2013})}\BibitemShut {NoStop}%
\bibitem [{\citenamefont {Yang}\ \emph {et~al.}(2014)\citenamefont {Yang},
  \citenamefont {Liu}, \citenamefont {Fu}, \citenamefont {Duan},\ and\
  \citenamefont {Liu}}]{Yang2014}%
  \BibitemOpen
  \bibfield  {author} {\bibinfo {author} {\bibfnamefont {G.}~\bibnamefont
  {Yang}}, \bibinfo {author} {\bibfnamefont {J.}~\bibnamefont {Liu}}, \bibinfo
  {author} {\bibfnamefont {L.}~\bibnamefont {Fu}}, \bibinfo {author}
  {\bibfnamefont {W.}~\bibnamefont {Duan}},\ and\ \bibinfo {author}
  {\bibfnamefont {C.}~\bibnamefont {Liu}},\ }\href
  {https://doi.org/10.1103/PhysRevB.89.085312} {\bibfield  {journal} {\bibinfo
  {journal} {Phys. Rev. B}\ }\textbf {\bibinfo {volume} {89}},\ \bibinfo
  {pages} {085312} (\bibinfo {year} {2014})}\BibitemShut {NoStop}%
\bibitem [{\citenamefont {Liu}\ \emph {et~al.}(2016)\citenamefont {Liu},
  \citenamefont {Zhou}, \citenamefont {Yao},\ and\ \citenamefont
  {Zhang}}]{Liu2016bb}%
  \BibitemOpen
  \bibfield  {author} {\bibinfo {author} {\bibfnamefont {C.-C.}\ \bibnamefont
  {Liu}}, \bibinfo {author} {\bibfnamefont {J.-J.}\ \bibnamefont {Zhou}},
  \bibinfo {author} {\bibfnamefont {Y.}~\bibnamefont {Yao}},\ and\ \bibinfo
  {author} {\bibfnamefont {F.}~\bibnamefont {Zhang}},\ }\href
  {https://doi.org/10.1103/PhysRevLett.116.066801} {\bibfield  {journal}
  {\bibinfo  {journal} {Phys. Rev. Lett.}\ }\textbf {\bibinfo {volume} {116}},\
  \bibinfo {pages} {066801} (\bibinfo {year} {2016})}\BibitemShut {NoStop}%
\bibitem [{\citenamefont {Zhang}\ \emph {et~al.}(2017)\citenamefont {Zhang},
  \citenamefont {Wang}, \citenamefont {Yu}, \citenamefont {Liu}, \citenamefont
  {Liang}, \citenamefont {Huang}, \citenamefont {Nie}, \citenamefont {Sun},
  \citenamefont {Zhang}, \citenamefont {Shen}, \citenamefont {Liu},
  \citenamefont {Weng}, \citenamefont {Zhao}, \citenamefont {Chen},
  \citenamefont {Jia}, \citenamefont {Hu}, \citenamefont {Ding}, \citenamefont
  {Zhao}, \citenamefont {Gao}, \citenamefont {Li}, \citenamefont {He},
  \citenamefont {Zhao}, \citenamefont {Zhang}, \citenamefont {Zhang},
  \citenamefont {Yang}, \citenamefont {Wang}, \citenamefont {Peng},
  \citenamefont {Dai}, \citenamefont {Fang}, \citenamefont {Xu}, \citenamefont
  {Chen},\ and\ \citenamefont {Zhou}}]{Zhang2017d}%
  \BibitemOpen
  \bibfield  {author} {\bibinfo {author} {\bibfnamefont {Y.}~\bibnamefont
  {Zhang}}, \bibinfo {author} {\bibfnamefont {C.}~\bibnamefont {Wang}},
  \bibinfo {author} {\bibfnamefont {L.}~\bibnamefont {Yu}}, \bibinfo {author}
  {\bibfnamefont {G.}~\bibnamefont {Liu}}, \bibinfo {author} {\bibfnamefont
  {A.}~\bibnamefont {Liang}}, \bibinfo {author} {\bibfnamefont
  {J.}~\bibnamefont {Huang}}, \bibinfo {author} {\bibfnamefont
  {S.}~\bibnamefont {Nie}}, \bibinfo {author} {\bibfnamefont {X.}~\bibnamefont
  {Sun}}, \bibinfo {author} {\bibfnamefont {Y.}~\bibnamefont {Zhang}}, \bibinfo
  {author} {\bibfnamefont {B.}~\bibnamefont {Shen}}, \bibinfo {author}
  {\bibfnamefont {J.}~\bibnamefont {Liu}}, \bibinfo {author} {\bibfnamefont
  {H.}~\bibnamefont {Weng}}, \bibinfo {author} {\bibfnamefont {L.}~\bibnamefont
  {Zhao}}, \bibinfo {author} {\bibfnamefont {G.}~\bibnamefont {Chen}}, \bibinfo
  {author} {\bibfnamefont {X.}~\bibnamefont {Jia}}, \bibinfo {author}
  {\bibfnamefont {C.}~\bibnamefont {Hu}}, \bibinfo {author} {\bibfnamefont
  {Y.}~\bibnamefont {Ding}}, \bibinfo {author} {\bibfnamefont {W.}~\bibnamefont
  {Zhao}}, \bibinfo {author} {\bibfnamefont {Q.}~\bibnamefont {Gao}}, \bibinfo
  {author} {\bibfnamefont {C.}~\bibnamefont {Li}}, \bibinfo {author}
  {\bibfnamefont {S.}~\bibnamefont {He}}, \bibinfo {author} {\bibfnamefont
  {L.}~\bibnamefont {Zhao}}, \bibinfo {author} {\bibfnamefont {F.}~\bibnamefont
  {Zhang}}, \bibinfo {author} {\bibfnamefont {S.}~\bibnamefont {Zhang}},
  \bibinfo {author} {\bibfnamefont {F.}~\bibnamefont {Yang}}, \bibinfo {author}
  {\bibfnamefont {Z.}~\bibnamefont {Wang}}, \bibinfo {author} {\bibfnamefont
  {Q.}~\bibnamefont {Peng}}, \bibinfo {author} {\bibfnamefont {X.}~\bibnamefont
  {Dai}}, \bibinfo {author} {\bibfnamefont {Z.}~\bibnamefont {Fang}}, \bibinfo
  {author} {\bibfnamefont {Z.}~\bibnamefont {Xu}}, \bibinfo {author}
  {\bibfnamefont {C.}~\bibnamefont {Chen}},\ and\ \bibinfo {author}
  {\bibfnamefont {X.~J.}\ \bibnamefont {Zhou}},\ }\href
  {https://doi.org/10.1038/ncomms15512} {\bibfield  {journal} {\bibinfo
  {journal} {Nat. Commun.}\ }\textbf {\bibinfo {volume} {8}},\ \bibinfo {pages}
  {15512} (\bibinfo {year} {2017})}\BibitemShut {NoStop}%
\bibitem [{\citenamefont {Liu}\ \emph {et~al.}(2018)\citenamefont {Liu},
  \citenamefont {Wang}, \citenamefont {Chen}, \citenamefont {Xu}, \citenamefont
  {Jiang}, \citenamefont {Yang}, \citenamefont {Yang}, \citenamefont {Lv},
  \citenamefont {Zhou}, \citenamefont {Chen}, \citenamefont {Yao},
  \citenamefont {Lu}, \citenamefont {Chen}, \citenamefont {Felser},
  \citenamefont {Yan}, \citenamefont {Liu},\ and\ \citenamefont
  {Chen}}]{Liu2018h}%
  \BibitemOpen
  \bibfield  {author} {\bibinfo {author} {\bibfnamefont {S.}~\bibnamefont
  {Liu}}, \bibinfo {author} {\bibfnamefont {M.~X.}\ \bibnamefont {Wang}},
  \bibinfo {author} {\bibfnamefont {C.}~\bibnamefont {Chen}}, \bibinfo {author}
  {\bibfnamefont {X.}~\bibnamefont {Xu}}, \bibinfo {author} {\bibfnamefont
  {J.}~\bibnamefont {Jiang}}, \bibinfo {author} {\bibfnamefont {L.~X.}\
  \bibnamefont {Yang}}, \bibinfo {author} {\bibfnamefont {H.~F.}\ \bibnamefont
  {Yang}}, \bibinfo {author} {\bibfnamefont {Y.~Y.}\ \bibnamefont {Lv}},
  \bibinfo {author} {\bibfnamefont {J.}~\bibnamefont {Zhou}}, \bibinfo {author}
  {\bibfnamefont {Y.~B.}\ \bibnamefont {Chen}}, \bibinfo {author}
  {\bibfnamefont {S.~H.}\ \bibnamefont {Yao}}, \bibinfo {author} {\bibfnamefont
  {M.~H.}\ \bibnamefont {Lu}}, \bibinfo {author} {\bibfnamefont {Y.~F.}\
  \bibnamefont {Chen}}, \bibinfo {author} {\bibfnamefont {C.}~\bibnamefont
  {Felser}}, \bibinfo {author} {\bibfnamefont {B.~H.}\ \bibnamefont {Yan}},
  \bibinfo {author} {\bibfnamefont {Z.~K.}\ \bibnamefont {Liu}},\ and\ \bibinfo
  {author} {\bibfnamefont {Y.~L.}\ \bibnamefont {Chen}},\ }\href
  {https://doi.org/10.1063/1.5050847} {\bibfield  {journal} {\bibinfo
  {journal} {APL Mater.}\ }\textbf {\bibinfo {volume} {6}},\ \bibinfo {pages}
  {121111} (\bibinfo {year} {2018})}\BibitemShut {NoStop}%
\bibitem [{\citenamefont {Noguchi}\ \emph {et~al.}(2019)\citenamefont
  {Noguchi}, \citenamefont {Takahashi}, \citenamefont {Kuroda}, \citenamefont
  {Ochi}, \citenamefont {Shirasawa}, \citenamefont {Sakano}, \citenamefont
  {Bareille}, \citenamefont {Nakayama}, \citenamefont {Watson}, \citenamefont
  {Yaji}, \citenamefont {Harasawa}, \citenamefont {Iwasawa}, \citenamefont
  {Dudin}, \citenamefont {Kim}, \citenamefont {Hoesch}, \citenamefont
  {Kandyba}, \citenamefont {Giampietri}, \citenamefont {Barinov}, \citenamefont
  {Shin}, \citenamefont {Arita}, \citenamefont {Sasagawa},\ and\ \citenamefont
  {Kondo}}]{Noguchi2019}%
  \BibitemOpen
  \bibfield  {author} {\bibinfo {author} {\bibfnamefont {R.}~\bibnamefont
  {Noguchi}}, \bibinfo {author} {\bibfnamefont {T.}~\bibnamefont {Takahashi}},
  \bibinfo {author} {\bibfnamefont {K.}~\bibnamefont {Kuroda}}, \bibinfo
  {author} {\bibfnamefont {M.}~\bibnamefont {Ochi}}, \bibinfo {author}
  {\bibfnamefont {T.}~\bibnamefont {Shirasawa}}, \bibinfo {author}
  {\bibfnamefont {M.}~\bibnamefont {Sakano}}, \bibinfo {author} {\bibfnamefont
  {C.}~\bibnamefont {Bareille}}, \bibinfo {author} {\bibfnamefont
  {M.}~\bibnamefont {Nakayama}}, \bibinfo {author} {\bibfnamefont {M.~D.}\
  \bibnamefont {Watson}}, \bibinfo {author} {\bibfnamefont {K.}~\bibnamefont
  {Yaji}}, \bibinfo {author} {\bibfnamefont {A.}~\bibnamefont {Harasawa}},
  \bibinfo {author} {\bibfnamefont {H.}~\bibnamefont {Iwasawa}}, \bibinfo
  {author} {\bibfnamefont {P.}~\bibnamefont {Dudin}}, \bibinfo {author}
  {\bibfnamefont {T.~K.}\ \bibnamefont {Kim}}, \bibinfo {author} {\bibfnamefont
  {M.}~\bibnamefont {Hoesch}}, \bibinfo {author} {\bibfnamefont
  {V.}~\bibnamefont {Kandyba}}, \bibinfo {author} {\bibfnamefont
  {A.}~\bibnamefont {Giampietri}}, \bibinfo {author} {\bibfnamefont
  {A.}~\bibnamefont {Barinov}}, \bibinfo {author} {\bibfnamefont
  {S.}~\bibnamefont {Shin}}, \bibinfo {author} {\bibfnamefont {R.}~\bibnamefont
  {Arita}}, \bibinfo {author} {\bibfnamefont {T.}~\bibnamefont {Sasagawa}},\
  and\ \bibinfo {author} {\bibfnamefont {T.}~\bibnamefont {Kondo}},\ }\href
  {https://doi.org/10.1038/s41586-019-0927-7} {\bibfield  {journal} {\bibinfo
  {journal} {Nature}\ }\textbf {\bibinfo {volume} {566}},\ \bibinfo {pages}
  {518} (\bibinfo {year} {2019})}\BibitemShut {NoStop}%
\bibitem [{\citenamefont {Huang}\ \emph {et~al.}(2021)\citenamefont {Huang},
  \citenamefont {Li}, \citenamefont {Yoon}, \citenamefont {Oh}, \citenamefont
  {Wu}, \citenamefont {Liu}, \citenamefont {Dhale}, \citenamefont {Zhou},
  \citenamefont {Guo}, \citenamefont {Zhang}, \citenamefont {Hashimoto},
  \citenamefont {Lu}, \citenamefont {Denlinger}, \citenamefont {Wang},
  \citenamefont {Lau}, \citenamefont {Birgeneau}, \citenamefont {Zhang},
  \citenamefont {Lv},\ and\ \citenamefont {Yi}}]{Huang2021}%
  \BibitemOpen
  \bibfield  {author} {\bibinfo {author} {\bibfnamefont {J.}~\bibnamefont
  {Huang}}, \bibinfo {author} {\bibfnamefont {S.}~\bibnamefont {Li}}, \bibinfo
  {author} {\bibfnamefont {C.}~\bibnamefont {Yoon}}, \bibinfo {author}
  {\bibfnamefont {J.~S.}\ \bibnamefont {Oh}}, \bibinfo {author} {\bibfnamefont
  {H.}~\bibnamefont {Wu}}, \bibinfo {author} {\bibfnamefont {X.}~\bibnamefont
  {Liu}}, \bibinfo {author} {\bibfnamefont {N.}~\bibnamefont {Dhale}}, \bibinfo
  {author} {\bibfnamefont {Y.-F.}\ \bibnamefont {Zhou}}, \bibinfo {author}
  {\bibfnamefont {Y.}~\bibnamefont {Guo}}, \bibinfo {author} {\bibfnamefont
  {Y.}~\bibnamefont {Zhang}}, \bibinfo {author} {\bibfnamefont
  {M.}~\bibnamefont {Hashimoto}}, \bibinfo {author} {\bibfnamefont
  {D.}~\bibnamefont {Lu}}, \bibinfo {author} {\bibfnamefont {J.}~\bibnamefont
  {Denlinger}}, \bibinfo {author} {\bibfnamefont {X.}~\bibnamefont {Wang}},
  \bibinfo {author} {\bibfnamefont {C.~N.}\ \bibnamefont {Lau}}, \bibinfo
  {author} {\bibfnamefont {R.~J.}\ \bibnamefont {Birgeneau}}, \bibinfo {author}
  {\bibfnamefont {F.}~\bibnamefont {Zhang}}, \bibinfo {author} {\bibfnamefont
  {B.}~\bibnamefont {Lv}},\ and\ \bibinfo {author} {\bibfnamefont
  {M.}~\bibnamefont {Yi}},\ }\href {https://doi.org/10.1103/PhysRevX.11.031042}
  {\bibfield  {journal} {\bibinfo  {journal} {Phys. Rev. X}\ }\textbf {\bibinfo
  {volume} {11}},\ \bibinfo {pages} {031042} (\bibinfo {year}
  {2021})}\BibitemShut {NoStop}%
\bibitem [{\citenamefont {Lee}\ \emph {et~al.}(2021)\citenamefont {Lee},
  \citenamefont {Lange}, \citenamefont {Wang}, \citenamefont {Kuthanazhi},
  \citenamefont {Trevisan}, \citenamefont {Jo}, \citenamefont {Schrunk},
  \citenamefont {Orth}, \citenamefont {Slager}, \citenamefont {Canfield},\ and\
  \citenamefont {Kaminski}}]{Lee2021}%
  \BibitemOpen
  \bibfield  {author} {\bibinfo {author} {\bibfnamefont {K.}~\bibnamefont
  {Lee}}, \bibinfo {author} {\bibfnamefont {G.~F.}\ \bibnamefont {Lange}},
  \bibinfo {author} {\bibfnamefont {L.-L.}\ \bibnamefont {Wang}}, \bibinfo
  {author} {\bibfnamefont {B.}~\bibnamefont {Kuthanazhi}}, \bibinfo {author}
  {\bibfnamefont {T.~V.}\ \bibnamefont {Trevisan}}, \bibinfo {author}
  {\bibfnamefont {N.~H.}\ \bibnamefont {Jo}}, \bibinfo {author} {\bibfnamefont
  {B.}~\bibnamefont {Schrunk}}, \bibinfo {author} {\bibfnamefont {P.~P.}\
  \bibnamefont {Orth}}, \bibinfo {author} {\bibfnamefont {R.-J.}\ \bibnamefont
  {Slager}}, \bibinfo {author} {\bibfnamefont {P.~C.}\ \bibnamefont
  {Canfield}},\ and\ \bibinfo {author} {\bibfnamefont {A.}~\bibnamefont
  {Kaminski}},\ }\href {https://doi.org/10.1038/s41467-021-22136-w} {\bibfield
  {journal} {\bibinfo  {journal} {Nat. Commun.}\ }\textbf {\bibinfo {volume}
  {12}},\ \bibinfo {pages} {1855} (\bibinfo {year} {2021})}\BibitemShut
  {NoStop}%
\bibitem [{\citenamefont {Ando}(2013)}]{Ando2013}%
  \BibitemOpen
  \bibfield  {author} {\bibinfo {author} {\bibfnamefont {Y.}~\bibnamefont
  {Ando}},\ }\href {https://doi.org/10.7566/JPSJ.82.102001} {\bibfield
  {journal} {\bibinfo  {journal} {J. Phys. Soc. Japan}\ }\textbf {\bibinfo
  {volume} {82}},\ \bibinfo {pages} {102001} (\bibinfo {year}
  {2013})}\BibitemShut {NoStop}%
\bibitem [{\citenamefont {Zhou}\ \emph {et~al.}(2014)\citenamefont {Zhou},
  \citenamefont {Feng}, \citenamefont {Liu}, \citenamefont {Guan},\ and\
  \citenamefont {Yao}}]{Zhou2014b}%
  \BibitemOpen
  \bibfield  {author} {\bibinfo {author} {\bibfnamefont {J.-J.}\ \bibnamefont
  {Zhou}}, \bibinfo {author} {\bibfnamefont {W.}~\bibnamefont {Feng}}, \bibinfo
  {author} {\bibfnamefont {C.-C.}\ \bibnamefont {Liu}}, \bibinfo {author}
  {\bibfnamefont {S.}~\bibnamefont {Guan}},\ and\ \bibinfo {author}
  {\bibfnamefont {Y.}~\bibnamefont {Yao}},\ }\href
  {https://doi.org/10.1021/nl501907g} {\bibfield  {journal} {\bibinfo
  {journal} {Nano Lett.}\ }\textbf {\bibinfo {volume} {14}},\ \bibinfo {pages}
  {4767} (\bibinfo {year} {2014})}\BibitemShut {NoStop}%
\bibitem [{\citenamefont {Zhou}\ \emph {et~al.}(2015)\citenamefont {Zhou},
  \citenamefont {Feng}, \citenamefont {Liu},\ and\ \citenamefont
  {Yao}}]{Zhou2015a}%
  \BibitemOpen
  \bibfield  {author} {\bibinfo {author} {\bibfnamefont {J.-J.}\ \bibnamefont
  {Zhou}}, \bibinfo {author} {\bibfnamefont {W.}~\bibnamefont {Feng}}, \bibinfo
  {author} {\bibfnamefont {G.-B.}\ \bibnamefont {Liu}},\ and\ \bibinfo {author}
  {\bibfnamefont {Y.}~\bibnamefont {Yao}},\ }\href
  {https://doi.org/10.1088/1367-2630/17/1/015004} {\bibfield  {journal}
  {\bibinfo  {journal} {New J. Phys.}\ }\textbf {\bibinfo {volume} {17}},\
  \bibinfo {pages} {015004} (\bibinfo {year} {2015})}\BibitemShut {NoStop}%
\bibitem [{\citenamefont {Aut{\`{e}}s}\ \emph {et~al.}(2016)\citenamefont
  {Aut{\`{e}}s}, \citenamefont {Isaeva}, \citenamefont {Moreschini},
  \citenamefont {Johannsen}, \citenamefont {Pisoni}, \citenamefont {Mori},
  \citenamefont {Zhang}, \citenamefont {Filatova}, \citenamefont {Kuznetsov},
  \citenamefont {Forr{\'{o}}}, \citenamefont {{Van den Broek}}, \citenamefont
  {Kim}, \citenamefont {Kim}, \citenamefont {Lanzara}, \citenamefont
  {Denlinger}, \citenamefont {Rotenberg}, \citenamefont {Bostwick},
  \citenamefont {Grioni},\ and\ \citenamefont {Yazyev}}]{Autes2015b}%
  \BibitemOpen
  \bibfield  {author} {\bibinfo {author} {\bibfnamefont {G.}~\bibnamefont
  {Aut{\`{e}}s}}, \bibinfo {author} {\bibfnamefont {A.}~\bibnamefont {Isaeva}},
  \bibinfo {author} {\bibfnamefont {L.}~\bibnamefont {Moreschini}}, \bibinfo
  {author} {\bibfnamefont {J.~C.}\ \bibnamefont {Johannsen}}, \bibinfo {author}
  {\bibfnamefont {A.}~\bibnamefont {Pisoni}}, \bibinfo {author} {\bibfnamefont
  {R.}~\bibnamefont {Mori}}, \bibinfo {author} {\bibfnamefont {W.}~\bibnamefont
  {Zhang}}, \bibinfo {author} {\bibfnamefont {T.~G.}\ \bibnamefont {Filatova}},
  \bibinfo {author} {\bibfnamefont {A.~N.}\ \bibnamefont {Kuznetsov}}, \bibinfo
  {author} {\bibfnamefont {L.}~\bibnamefont {Forr{\'{o}}}}, \bibinfo {author}
  {\bibfnamefont {W.}~\bibnamefont {{Van den Broek}}}, \bibinfo {author}
  {\bibfnamefont {Y.}~\bibnamefont {Kim}}, \bibinfo {author} {\bibfnamefont
  {K.~S.}\ \bibnamefont {Kim}}, \bibinfo {author} {\bibfnamefont
  {A.}~\bibnamefont {Lanzara}}, \bibinfo {author} {\bibfnamefont {J.~D.}\
  \bibnamefont {Denlinger}}, \bibinfo {author} {\bibfnamefont {E.}~\bibnamefont
  {Rotenberg}}, \bibinfo {author} {\bibfnamefont {A.}~\bibnamefont {Bostwick}},
  \bibinfo {author} {\bibfnamefont {M.}~\bibnamefont {Grioni}},\ and\ \bibinfo
  {author} {\bibfnamefont {O.~V.}\ \bibnamefont {Yazyev}},\ }\href
  {https://doi.org/10.1038/nmat4488} {\bibfield  {journal} {\bibinfo  {journal}
  {Nat. Mater.}\ }\textbf {\bibinfo {volume} {15}},\ \bibinfo {pages} {154}
  (\bibinfo {year} {2016})}\BibitemShut {NoStop}%
\bibitem [{\citenamefont {Pisoni}\ \emph {et~al.}(2017)\citenamefont {Pisoni},
  \citenamefont {Gaal}, \citenamefont {Zeugner}, \citenamefont {Falkowski},
  \citenamefont {Isaeva}, \citenamefont {Huppertz}, \citenamefont {Autes},
  \citenamefont {Yazyev},\ and\ \citenamefont {Forro}}]{InsulatorPisoni}%
  \BibitemOpen
  \bibfield  {author} {\bibinfo {author} {\bibfnamefont {A.}~\bibnamefont
  {Pisoni}}, \bibinfo {author} {\bibfnamefont {R.}~\bibnamefont {Gaal}},
  \bibinfo {author} {\bibfnamefont {A.}~\bibnamefont {Zeugner}}, \bibinfo
  {author} {\bibfnamefont {V.}~\bibnamefont {Falkowski}}, \bibinfo {author}
  {\bibfnamefont {A.}~\bibnamefont {Isaeva}}, \bibinfo {author} {\bibfnamefont
  {H.}~\bibnamefont {Huppertz}}, \bibinfo {author} {\bibfnamefont
  {G.}~\bibnamefont {Autes}}, \bibinfo {author} {\bibfnamefont {O.~V.}\
  \bibnamefont {Yazyev}},\ and\ \bibinfo {author} {\bibfnamefont
  {L.}~\bibnamefont {Forro}},\ }\href
  {https://doi.org/10.1103/PhysRevB.95.235149} {\bibfield  {journal} {\bibinfo
  {journal} {Phys. Rev. B}\ }\textbf {\bibinfo {volume} {95}},\ \bibinfo
  {pages} {235149} (\bibinfo {year} {2017})}\BibitemShut {NoStop}%
\bibitem [{\citenamefont {Qi}\ \emph {et~al.}(2018)\citenamefont {Qi},
  \citenamefont {Shi}, \citenamefont {Werner}, \citenamefont {Naumov},
  \citenamefont {Schnelle}, \citenamefont {Wang}, \citenamefont {Rana},
  \citenamefont {Parkin}, \citenamefont {Medvedev}, \citenamefont {Yan},\ and\
  \citenamefont {Felser}}]{Qi2017}%
  \BibitemOpen
  \bibfield  {author} {\bibinfo {author} {\bibfnamefont {Y.}~\bibnamefont
  {Qi}}, \bibinfo {author} {\bibfnamefont {W.}~\bibnamefont {Shi}}, \bibinfo
  {author} {\bibfnamefont {P.}~\bibnamefont {Werner}}, \bibinfo {author}
  {\bibfnamefont {P.~G.}\ \bibnamefont {Naumov}}, \bibinfo {author}
  {\bibfnamefont {W.}~\bibnamefont {Schnelle}}, \bibinfo {author}
  {\bibfnamefont {L.}~\bibnamefont {Wang}}, \bibinfo {author} {\bibfnamefont
  {K.~G.}\ \bibnamefont {Rana}}, \bibinfo {author} {\bibfnamefont
  {S.}~\bibnamefont {Parkin}}, \bibinfo {author} {\bibfnamefont {S.~A.}\
  \bibnamefont {Medvedev}}, \bibinfo {author} {\bibfnamefont {B.}~\bibnamefont
  {Yan}},\ and\ \bibinfo {author} {\bibfnamefont {C.}~\bibnamefont {Felser}},\
  }\href {https://doi.org/10.1038/s41535-018-0078-3} {\bibfield  {journal}
  {\bibinfo  {journal} {npj Quantum Mater.}\ }\textbf {\bibinfo {volume} {3}},\
  \bibinfo {pages} {4} (\bibinfo {year} {2018})}\BibitemShut {NoStop}%
\bibitem [{\citenamefont {Wang}\ \emph {et~al.}(2018)\citenamefont {Wang},
  \citenamefont {Wu}, \citenamefont {Wang}, \citenamefont {Chen}, \citenamefont
  {Gao}, \citenamefont {Lu}, \citenamefont {Chen}, \citenamefont {Ding},
  \citenamefont {Wen},\ and\ \citenamefont {Sun}}]{Wang2018}%
  \BibitemOpen
  \bibfield  {author} {\bibinfo {author} {\bibfnamefont {X.}~\bibnamefont
  {Wang}}, \bibinfo {author} {\bibfnamefont {J.}~\bibnamefont {Wu}}, \bibinfo
  {author} {\bibfnamefont {J.}~\bibnamefont {Wang}}, \bibinfo {author}
  {\bibfnamefont {T.}~\bibnamefont {Chen}}, \bibinfo {author} {\bibfnamefont
  {H.}~\bibnamefont {Gao}}, \bibinfo {author} {\bibfnamefont {P.}~\bibnamefont
  {Lu}}, \bibinfo {author} {\bibfnamefont {Q.}~\bibnamefont {Chen}}, \bibinfo
  {author} {\bibfnamefont {C.}~\bibnamefont {Ding}}, \bibinfo {author}
  {\bibfnamefont {J.}~\bibnamefont {Wen}},\ and\ \bibinfo {author}
  {\bibfnamefont {J.}~\bibnamefont {Sun}},\ }\href
  {https://doi.org/10.1103/PhysRevB.98.174112} {\bibfield  {journal} {\bibinfo
  {journal} {Phys. Rev. B}\ }\textbf {\bibinfo {volume} {98}},\ \bibinfo
  {pages} {174112} (\bibinfo {year} {2018})}\BibitemShut {NoStop}%
\bibitem [{\citenamefont {Chen}\ \emph {et~al.}(2018)\citenamefont {Chen},
  \citenamefont {Ma}, \citenamefont {Li}, \citenamefont {Du}, \citenamefont
  {Xiong}, \citenamefont {He}, \citenamefont {Duan}, \citenamefont {Han},
  \citenamefont {Chen}, \citenamefont {Xiao},\ and\ \citenamefont
  {Yao}}]{Chen2018f}%
  \BibitemOpen
  \bibfield  {author} {\bibinfo {author} {\bibfnamefont {D.-Y.}\ \bibnamefont
  {Chen}}, \bibinfo {author} {\bibfnamefont {D.-s.}\ \bibnamefont {Ma}},
  \bibinfo {author} {\bibfnamefont {Y.}~\bibnamefont {Li}}, \bibinfo {author}
  {\bibfnamefont {Z.~Z.}\ \bibnamefont {Du}}, \bibinfo {author} {\bibfnamefont
  {X.}~\bibnamefont {Xiong}}, \bibinfo {author} {\bibfnamefont
  {Y.}~\bibnamefont {He}}, \bibinfo {author} {\bibfnamefont {J.}~\bibnamefont
  {Duan}}, \bibinfo {author} {\bibfnamefont {J.}~\bibnamefont {Han}}, \bibinfo
  {author} {\bibfnamefont {D.}~\bibnamefont {Chen}}, \bibinfo {author}
  {\bibfnamefont {W.}~\bibnamefont {Xiao}},\ and\ \bibinfo {author}
  {\bibfnamefont {Y.}~\bibnamefont {Yao}},\ }\href
  {https://doi.org/10.1103/PhysRevMaterials.2.114408} {\bibfield  {journal}
  {\bibinfo  {journal} {Phys. Rev. Mater.}\ }\textbf {\bibinfo {volume} {2}},\
  \bibinfo {pages} {114408} (\bibinfo {year} {2018})}\BibitemShut {NoStop}%
\bibitem [{\citenamefont {Deng}\ \emph {et~al.}(2019)\citenamefont {Deng},
  \citenamefont {Song}, \citenamefont {Shao}, \citenamefont {Li}, \citenamefont
  {Xie}, \citenamefont {Chen},\ and\ \citenamefont {Ma}}]{Deng2019a}%
  \BibitemOpen
  \bibfield  {author} {\bibinfo {author} {\bibfnamefont {S.}~\bibnamefont
  {Deng}}, \bibinfo {author} {\bibfnamefont {X.}~\bibnamefont {Song}}, \bibinfo
  {author} {\bibfnamefont {X.}~\bibnamefont {Shao}}, \bibinfo {author}
  {\bibfnamefont {Q.}~\bibnamefont {Li}}, \bibinfo {author} {\bibfnamefont
  {Y.}~\bibnamefont {Xie}}, \bibinfo {author} {\bibfnamefont {C.}~\bibnamefont
  {Chen}},\ and\ \bibinfo {author} {\bibfnamefont {Y.}~\bibnamefont {Ma}},\
  }\href {https://doi.org/10.1103/PhysRevB.100.224108} {\bibfield  {journal}
  {\bibinfo  {journal} {Phys. Rev. B}\ }\textbf {\bibinfo {volume} {100}},\
  \bibinfo {pages} {224108} (\bibinfo {year} {2019})}\BibitemShut {NoStop}%
\bibitem [{\citenamefont {Li}\ \emph {et~al.}(2019)\citenamefont {Li},
  \citenamefont {Chen}, \citenamefont {Jin}, \citenamefont {Ma}, \citenamefont
  {Ge}, \citenamefont {Sun}, \citenamefont {Guo}, \citenamefont {Sun},
  \citenamefont {Han}, \citenamefont {Xiao}, \citenamefont {Duan},
  \citenamefont {Wang}, \citenamefont {Liu}, \citenamefont {Zou}, \citenamefont
  {Cheng}, \citenamefont {Jin}, \citenamefont {Zhou}, \citenamefont
  {Goodenough}, \citenamefont {Zhu},\ and\ \citenamefont {Yao}}]{Li2019d}%
  \BibitemOpen
  \bibfield  {author} {\bibinfo {author} {\bibfnamefont {X.}~\bibnamefont
  {Li}}, \bibinfo {author} {\bibfnamefont {D.}~\bibnamefont {Chen}}, \bibinfo
  {author} {\bibfnamefont {M.}~\bibnamefont {Jin}}, \bibinfo {author}
  {\bibfnamefont {D.}~\bibnamefont {Ma}}, \bibinfo {author} {\bibfnamefont
  {Y.}~\bibnamefont {Ge}}, \bibinfo {author} {\bibfnamefont {J.}~\bibnamefont
  {Sun}}, \bibinfo {author} {\bibfnamefont {W.}~\bibnamefont {Guo}}, \bibinfo
  {author} {\bibfnamefont {H.}~\bibnamefont {Sun}}, \bibinfo {author}
  {\bibfnamefont {J.}~\bibnamefont {Han}}, \bibinfo {author} {\bibfnamefont
  {W.}~\bibnamefont {Xiao}}, \bibinfo {author} {\bibfnamefont {J.}~\bibnamefont
  {Duan}}, \bibinfo {author} {\bibfnamefont {Q.}~\bibnamefont {Wang}}, \bibinfo
  {author} {\bibfnamefont {C.-C.}\ \bibnamefont {Liu}}, \bibinfo {author}
  {\bibfnamefont {R.}~\bibnamefont {Zou}}, \bibinfo {author} {\bibfnamefont
  {J.}~\bibnamefont {Cheng}}, \bibinfo {author} {\bibfnamefont
  {C.}~\bibnamefont {Jin}}, \bibinfo {author} {\bibfnamefont {J.}~\bibnamefont
  {Zhou}}, \bibinfo {author} {\bibfnamefont {J.~B.}\ \bibnamefont
  {Goodenough}}, \bibinfo {author} {\bibfnamefont {J.}~\bibnamefont {Zhu}},\
  and\ \bibinfo {author} {\bibfnamefont {Y.}~\bibnamefont {Yao}},\ }\href
  {https://doi.org/10.1073/pnas.1909276116} {\bibfield  {journal} {\bibinfo
  {journal} {Proc. Natl. Acad. Sci.}\ }\textbf {\bibinfo {volume} {116}},\
  \bibinfo {pages} {17696} (\bibinfo {year} {2019})}\BibitemShut {NoStop}%
\bibitem [{\citenamefont {Hsu}\ \emph {et~al.}(2019)\citenamefont {Hsu},
  \citenamefont {Zhou}, \citenamefont {Ma}, \citenamefont {Gedik},
  \citenamefont {Bansil}, \citenamefont {Pereira}, \citenamefont {Lin},
  \citenamefont {Fu}, \citenamefont {Xu},\ and\ \citenamefont
  {Chang}}]{Hsu2019b}%
  \BibitemOpen
  \bibfield  {author} {\bibinfo {author} {\bibfnamefont {C.-H.}\ \bibnamefont
  {Hsu}}, \bibinfo {author} {\bibfnamefont {X.}~\bibnamefont {Zhou}}, \bibinfo
  {author} {\bibfnamefont {Q.}~\bibnamefont {Ma}}, \bibinfo {author}
  {\bibfnamefont {N.}~\bibnamefont {Gedik}}, \bibinfo {author} {\bibfnamefont
  {A.}~\bibnamefont {Bansil}}, \bibinfo {author} {\bibfnamefont {V.~M.}\
  \bibnamefont {Pereira}}, \bibinfo {author} {\bibfnamefont {H.}~\bibnamefont
  {Lin}}, \bibinfo {author} {\bibfnamefont {L.}~\bibnamefont {Fu}}, \bibinfo
  {author} {\bibfnamefont {S.-Y.}\ \bibnamefont {Xu}},\ and\ \bibinfo {author}
  {\bibfnamefont {T.-R.}\ \bibnamefont {Chang}},\ }\href
  {https://doi.org/10.1088/2053-1583/ab1607} {\bibfield  {journal} {\bibinfo
  {journal} {2D Mater.}\ }\textbf {\bibinfo {volume} {6}},\ \bibinfo {pages}
  {031004} (\bibinfo {year} {2019})}\BibitemShut {NoStop}%
\bibitem [{\citenamefont {Qiao}\ \emph {et~al.}(2021)\citenamefont {Qiao},
  \citenamefont {Xiong}, \citenamefont {Yang}, \citenamefont {Chen},
  \citenamefont {Li}, \citenamefont {Li}, \citenamefont {Peng}, \citenamefont
  {Xu}, \citenamefont {Han}, \citenamefont {Xiao},\ and\ \citenamefont
  {Yao}}]{Qiao2021}%
  \BibitemOpen
  \bibfield  {author} {\bibinfo {author} {\bibfnamefont {L.}~\bibnamefont
  {Qiao}}, \bibinfo {author} {\bibfnamefont {X.}~\bibnamefont {Xiong}},
  \bibinfo {author} {\bibfnamefont {H.}~\bibnamefont {Yang}}, \bibinfo {author}
  {\bibfnamefont {D.}~\bibnamefont {Chen}}, \bibinfo {author} {\bibfnamefont
  {Y.}~\bibnamefont {Li}}, \bibinfo {author} {\bibfnamefont {J.}~\bibnamefont
  {Li}}, \bibinfo {author} {\bibfnamefont {X.}~\bibnamefont {Peng}}, \bibinfo
  {author} {\bibfnamefont {Z.}~\bibnamefont {Xu}}, \bibinfo {author}
  {\bibfnamefont {J.}~\bibnamefont {Han}}, \bibinfo {author} {\bibfnamefont
  {W.}~\bibnamefont {Xiao}},\ and\ \bibinfo {author} {\bibfnamefont
  {Y.}~\bibnamefont {Yao}},\ }\href {https://doi.org/10.1021/acs.jpcc.1c06702}
  {\bibfield  {journal} {\bibinfo  {journal} {J. Phys. Chem. C}\ }\textbf
  {\bibinfo {volume} {125}},\ \bibinfo {pages} {22312} (\bibinfo {year}
  {2021})}\BibitemShut {NoStop}%
\bibitem [{\citenamefont {Noguchi}\ \emph {et~al.}(2021)\citenamefont
  {Noguchi}, \citenamefont {Kobayashi}, \citenamefont {Jiang}, \citenamefont
  {Kuroda}, \citenamefont {Takahashi}, \citenamefont {Xu}, \citenamefont {Lee},
  \citenamefont {Hirayama}, \citenamefont {Ochi}, \citenamefont {Shirasawa},
  \citenamefont {Zhang}, \citenamefont {Lin}, \citenamefont {Bareille},
  \citenamefont {Sakuragi}, \citenamefont {Tanaka}, \citenamefont {Kunisada},
  \citenamefont {Kurokawa}, \citenamefont {Yaji}, \citenamefont {Harasawa},
  \citenamefont {Kandyba}, \citenamefont {Giampietri}, \citenamefont {Barinov},
  \citenamefont {Kim}, \citenamefont {Cacho}, \citenamefont {Hashimoto},
  \citenamefont {Lu}, \citenamefont {Shin}, \citenamefont {Arita},
  \citenamefont {Lai}, \citenamefont {Sasagawa},\ and\ \citenamefont
  {Kondo}}]{Noguchi2021a}%
  \BibitemOpen
  \bibfield  {author} {\bibinfo {author} {\bibfnamefont {R.}~\bibnamefont
  {Noguchi}}, \bibinfo {author} {\bibfnamefont {M.}~\bibnamefont {Kobayashi}},
  \bibinfo {author} {\bibfnamefont {Z.}~\bibnamefont {Jiang}}, \bibinfo
  {author} {\bibfnamefont {K.}~\bibnamefont {Kuroda}}, \bibinfo {author}
  {\bibfnamefont {T.}~\bibnamefont {Takahashi}}, \bibinfo {author}
  {\bibfnamefont {Z.}~\bibnamefont {Xu}}, \bibinfo {author} {\bibfnamefont
  {D.}~\bibnamefont {Lee}}, \bibinfo {author} {\bibfnamefont {M.}~\bibnamefont
  {Hirayama}}, \bibinfo {author} {\bibfnamefont {M.}~\bibnamefont {Ochi}},
  \bibinfo {author} {\bibfnamefont {T.}~\bibnamefont {Shirasawa}}, \bibinfo
  {author} {\bibfnamefont {P.}~\bibnamefont {Zhang}}, \bibinfo {author}
  {\bibfnamefont {C.}~\bibnamefont {Lin}}, \bibinfo {author} {\bibfnamefont
  {C.}~\bibnamefont {Bareille}}, \bibinfo {author} {\bibfnamefont
  {S.}~\bibnamefont {Sakuragi}}, \bibinfo {author} {\bibfnamefont
  {H.}~\bibnamefont {Tanaka}}, \bibinfo {author} {\bibfnamefont
  {S.}~\bibnamefont {Kunisada}}, \bibinfo {author} {\bibfnamefont
  {K.}~\bibnamefont {Kurokawa}}, \bibinfo {author} {\bibfnamefont
  {K.}~\bibnamefont {Yaji}}, \bibinfo {author} {\bibfnamefont {A.}~\bibnamefont
  {Harasawa}}, \bibinfo {author} {\bibfnamefont {V.}~\bibnamefont {Kandyba}},
  \bibinfo {author} {\bibfnamefont {A.}~\bibnamefont {Giampietri}}, \bibinfo
  {author} {\bibfnamefont {A.}~\bibnamefont {Barinov}}, \bibinfo {author}
  {\bibfnamefont {T.~K.}\ \bibnamefont {Kim}}, \bibinfo {author} {\bibfnamefont
  {C.}~\bibnamefont {Cacho}}, \bibinfo {author} {\bibfnamefont
  {M.}~\bibnamefont {Hashimoto}}, \bibinfo {author} {\bibfnamefont
  {D.}~\bibnamefont {Lu}}, \bibinfo {author} {\bibfnamefont {S.}~\bibnamefont
  {Shin}}, \bibinfo {author} {\bibfnamefont {R.}~\bibnamefont {Arita}},
  \bibinfo {author} {\bibfnamefont {K.}~\bibnamefont {Lai}}, \bibinfo {author}
  {\bibfnamefont {T.}~\bibnamefont {Sasagawa}},\ and\ \bibinfo {author}
  {\bibfnamefont {T.}~\bibnamefont {Kondo}},\ }\href
  {https://doi.org/10.1038/s41563-020-00871-7} {\bibfield  {journal} {\bibinfo
  {journal} {Nat. Mater.}\ }\textbf {\bibinfo {volume} {20}},\ \bibinfo {pages}
  {473} (\bibinfo {year} {2021})}\BibitemShut {NoStop}%
\bibitem [{\citenamefont {Wang}\ \emph {et~al.}(2021)\citenamefont {Wang},
  \citenamefont {Tang}, \citenamefont {Wang}, \citenamefont {Zhu},
  \citenamefont {Cho}, \citenamefont {Wang}, \citenamefont {Du}, \citenamefont
  {Shao},\ and\ \citenamefont {Zhang}}]{Wang2021}%
  \BibitemOpen
  \bibfield  {author} {\bibinfo {author} {\bibfnamefont {P.}~\bibnamefont
  {Wang}}, \bibinfo {author} {\bibfnamefont {F.}~\bibnamefont {Tang}}, \bibinfo
  {author} {\bibfnamefont {P.}~\bibnamefont {Wang}}, \bibinfo {author}
  {\bibfnamefont {H.}~\bibnamefont {Zhu}}, \bibinfo {author} {\bibfnamefont
  {C.-w.}\ \bibnamefont {Cho}}, \bibinfo {author} {\bibfnamefont
  {J.}~\bibnamefont {Wang}}, \bibinfo {author} {\bibfnamefont {X.}~\bibnamefont
  {Du}}, \bibinfo {author} {\bibfnamefont {Y.}~\bibnamefont {Shao}},\ and\
  \bibinfo {author} {\bibfnamefont {L.}~\bibnamefont {Zhang}},\ }\href
  {https://doi.org/10.1103/PhysRevB.103.155201} {\bibfield  {journal} {\bibinfo
   {journal} {Phys. Rev. B}\ }\textbf {\bibinfo {volume} {103}},\ \bibinfo
  {pages} {155201} (\bibinfo {year} {2021})}\BibitemShut {NoStop}%
\bibitem [{\citenamefont {Zhuang}\ \emph {et~al.}(2021)\citenamefont {Zhuang},
  \citenamefont {Li}, \citenamefont {Liu}, \citenamefont {Mu}, \citenamefont
  {Yang}, \citenamefont {Liu}, \citenamefont {Zhou}, \citenamefont {Hao},
  \citenamefont {Zhong},\ and\ \citenamefont {Du}}]{Zhuang2021}%
  \BibitemOpen
  \bibfield  {author} {\bibinfo {author} {\bibfnamefont {J.}~\bibnamefont
  {Zhuang}}, \bibinfo {author} {\bibfnamefont {J.}~\bibnamefont {Li}}, \bibinfo
  {author} {\bibfnamefont {Y.}~\bibnamefont {Liu}}, \bibinfo {author}
  {\bibfnamefont {D.}~\bibnamefont {Mu}}, \bibinfo {author} {\bibfnamefont
  {M.}~\bibnamefont {Yang}}, \bibinfo {author} {\bibfnamefont {Y.}~\bibnamefont
  {Liu}}, \bibinfo {author} {\bibfnamefont {W.}~\bibnamefont {Zhou}}, \bibinfo
  {author} {\bibfnamefont {W.}~\bibnamefont {Hao}}, \bibinfo {author}
  {\bibfnamefont {J.}~\bibnamefont {Zhong}},\ and\ \bibinfo {author}
  {\bibfnamefont {Y.}~\bibnamefont {Du}},\ }\href
  {https://doi.org/10.1021/acsnano.1c04928} {\bibfield  {journal} {\bibinfo
  {journal} {ACS Nano}\ }\textbf {\bibinfo {volume} {15}},\ \bibinfo {pages}
  {14850} (\bibinfo {year} {2021})}\BibitemShut {NoStop}%
\bibitem [{\citenamefont {Peng}\ \emph {et~al.}(2021)\citenamefont {Peng},
  \citenamefont {Zhang}, \citenamefont {Dong}, \citenamefont {Ma},
  \citenamefont {Chen}, \citenamefont {Li}, \citenamefont {Li}, \citenamefont
  {Han}, \citenamefont {Wang}, \citenamefont {Liu}, \citenamefont {Zhou},
  \citenamefont {Xiao},\ and\ \citenamefont {Yao}}]{Peng2021}%
  \BibitemOpen
  \bibfield  {author} {\bibinfo {author} {\bibfnamefont {X.}~\bibnamefont
  {Peng}}, \bibinfo {author} {\bibfnamefont {X.}~\bibnamefont {Zhang}},
  \bibinfo {author} {\bibfnamefont {X.}~\bibnamefont {Dong}}, \bibinfo {author}
  {\bibfnamefont {D.}~\bibnamefont {Ma}}, \bibinfo {author} {\bibfnamefont
  {D.}~\bibnamefont {Chen}}, \bibinfo {author} {\bibfnamefont {Y.}~\bibnamefont
  {Li}}, \bibinfo {author} {\bibfnamefont {J.}~\bibnamefont {Li}}, \bibinfo
  {author} {\bibfnamefont {J.}~\bibnamefont {Han}}, \bibinfo {author}
  {\bibfnamefont {Z.}~\bibnamefont {Wang}}, \bibinfo {author} {\bibfnamefont
  {C.-C.}\ \bibnamefont {Liu}}, \bibinfo {author} {\bibfnamefont
  {J.}~\bibnamefont {Zhou}}, \bibinfo {author} {\bibfnamefont {W.}~\bibnamefont
  {Xiao}},\ and\ \bibinfo {author} {\bibfnamefont {Y.}~\bibnamefont {Yao}},\
  }\href {https://doi.org/10.1021/acs.jpclett.1c02586} {\bibfield  {journal}
  {\bibinfo  {journal} {J. Phys. Chem. Lett.}\ }\textbf {\bibinfo {volume}
  {12}},\ \bibinfo {pages} {10465} (\bibinfo {year} {2021})}\BibitemShut
  {NoStop}%
\bibitem [{\citenamefont {Chen}\ \emph {et~al.}(2022)\citenamefont {Chen},
  \citenamefont {Ma}, \citenamefont {Duan}, \citenamefont {Chen}, \citenamefont
  {Liu}, \citenamefont {Han},\ and\ \citenamefont {Yao}}]{Chen2022}%
  \BibitemOpen
  \bibfield  {author} {\bibinfo {author} {\bibfnamefont {D.-Y.}\ \bibnamefont
  {Chen}}, \bibinfo {author} {\bibfnamefont {D.}~\bibnamefont {Ma}}, \bibinfo
  {author} {\bibfnamefont {J.}~\bibnamefont {Duan}}, \bibinfo {author}
  {\bibfnamefont {D.}~\bibnamefont {Chen}}, \bibinfo {author} {\bibfnamefont
  {H.}~\bibnamefont {Liu}}, \bibinfo {author} {\bibfnamefont {J.}~\bibnamefont
  {Han}},\ and\ \bibinfo {author} {\bibfnamefont {Y.}~\bibnamefont {Yao}},\
  }\href {https://doi.org/10.1103/PhysRevB.106.075206} {\bibfield  {journal}
  {\bibinfo  {journal} {Phys. Rev. B}\ }\textbf {\bibinfo {volume} {106}},\
  \bibinfo {pages} {075206} (\bibinfo {year} {2022})}\BibitemShut {NoStop}%
\bibitem [{\citenamefont {Shumiya}\ \emph {et~al.}(2022)\citenamefont
  {Shumiya}, \citenamefont {Hossain}, \citenamefont {Yin}, \citenamefont
  {Wang}, \citenamefont {Litskevich}, \citenamefont {Yoon}, \citenamefont {Li},
  \citenamefont {Yang}, \citenamefont {Jiang}, \citenamefont {Cheng},
  \citenamefont {Lin}, \citenamefont {Zhang}, \citenamefont {Cheng},
  \citenamefont {Cochran}, \citenamefont {Multer}, \citenamefont {Yang},
  \citenamefont {Casas}, \citenamefont {Chang}, \citenamefont {Neupert},
  \citenamefont {Yuan}, \citenamefont {Jia}, \citenamefont {Lin}, \citenamefont
  {Yao}, \citenamefont {Balicas}, \citenamefont {Zhang}, \citenamefont {Yao},\
  and\ \citenamefont {Hasan}}]{Shumiya2022}%
  \BibitemOpen
  \bibfield  {author} {\bibinfo {author} {\bibfnamefont {N.}~\bibnamefont
  {Shumiya}}, \bibinfo {author} {\bibfnamefont {M.~S.}\ \bibnamefont
  {Hossain}}, \bibinfo {author} {\bibfnamefont {J.-X.}\ \bibnamefont {Yin}},
  \bibinfo {author} {\bibfnamefont {Z.}~\bibnamefont {Wang}}, \bibinfo {author}
  {\bibfnamefont {M.}~\bibnamefont {Litskevich}}, \bibinfo {author}
  {\bibfnamefont {C.}~\bibnamefont {Yoon}}, \bibinfo {author} {\bibfnamefont
  {Y.}~\bibnamefont {Li}}, \bibinfo {author} {\bibfnamefont {Y.}~\bibnamefont
  {Yang}}, \bibinfo {author} {\bibfnamefont {Y.-X.}\ \bibnamefont {Jiang}},
  \bibinfo {author} {\bibfnamefont {G.}~\bibnamefont {Cheng}}, \bibinfo
  {author} {\bibfnamefont {Y.-C.}\ \bibnamefont {Lin}}, \bibinfo {author}
  {\bibfnamefont {Q.}~\bibnamefont {Zhang}}, \bibinfo {author} {\bibfnamefont
  {Z.-J.}\ \bibnamefont {Cheng}}, \bibinfo {author} {\bibfnamefont {T.~A.}\
  \bibnamefont {Cochran}}, \bibinfo {author} {\bibfnamefont {D.}~\bibnamefont
  {Multer}}, \bibinfo {author} {\bibfnamefont {X.~P.}\ \bibnamefont {Yang}},
  \bibinfo {author} {\bibfnamefont {B.}~\bibnamefont {Casas}}, \bibinfo
  {author} {\bibfnamefont {T.-R.}\ \bibnamefont {Chang}}, \bibinfo {author}
  {\bibfnamefont {T.}~\bibnamefont {Neupert}}, \bibinfo {author} {\bibfnamefont
  {Z.}~\bibnamefont {Yuan}}, \bibinfo {author} {\bibfnamefont {S.}~\bibnamefont
  {Jia}}, \bibinfo {author} {\bibfnamefont {H.}~\bibnamefont {Lin}}, \bibinfo
  {author} {\bibfnamefont {N.}~\bibnamefont {Yao}}, \bibinfo {author}
  {\bibfnamefont {L.}~\bibnamefont {Balicas}}, \bibinfo {author} {\bibfnamefont
  {F.}~\bibnamefont {Zhang}}, \bibinfo {author} {\bibfnamefont
  {Y.}~\bibnamefont {Yao}},\ and\ \bibinfo {author} {\bibfnamefont {M.~Z.}\
  \bibnamefont {Hasan}},\ }\href {https://doi.org/10.1038/s41563-022-01304-3}
  {\bibfield  {journal} {\bibinfo  {journal} {Nat. Mater.}\ }\textbf {\bibinfo
  {volume} {21}},\ \bibinfo {pages} {1111} (\bibinfo {year}
  {2022})}\BibitemShut {NoStop}%
\bibitem [{\citenamefont {Liu}\ \emph {et~al.}(2022)\citenamefont {Liu},
  \citenamefont {Chen}, \citenamefont {Zhang}, \citenamefont {Bockrath},
  \citenamefont {Lau}, \citenamefont {Zhou}, \citenamefont {Yoon},
  \citenamefont {Li}, \citenamefont {Liu}, \citenamefont {Dhale}, \citenamefont
  {Lv}, \citenamefont {Zhang}, \citenamefont {Watanabe}, \citenamefont
  {Taniguchi}, \citenamefont {Huang}, \citenamefont {Yi}, \citenamefont {Oh},\
  and\ \citenamefont {Birgeneau}}]{Liu2022}%
  \BibitemOpen
  \bibfield  {author} {\bibinfo {author} {\bibfnamefont {Y.}~\bibnamefont
  {Liu}}, \bibinfo {author} {\bibfnamefont {R.}~\bibnamefont {Chen}}, \bibinfo
  {author} {\bibfnamefont {Z.}~\bibnamefont {Zhang}}, \bibinfo {author}
  {\bibfnamefont {M.}~\bibnamefont {Bockrath}}, \bibinfo {author}
  {\bibfnamefont {C.~N.}\ \bibnamefont {Lau}}, \bibinfo {author} {\bibfnamefont
  {Y.-F.}\ \bibnamefont {Zhou}}, \bibinfo {author} {\bibfnamefont
  {C.}~\bibnamefont {Yoon}}, \bibinfo {author} {\bibfnamefont {S.}~\bibnamefont
  {Li}}, \bibinfo {author} {\bibfnamefont {X.}~\bibnamefont {Liu}}, \bibinfo
  {author} {\bibfnamefont {N.}~\bibnamefont {Dhale}}, \bibinfo {author}
  {\bibfnamefont {B.}~\bibnamefont {Lv}}, \bibinfo {author} {\bibfnamefont
  {F.}~\bibnamefont {Zhang}}, \bibinfo {author} {\bibfnamefont
  {K.}~\bibnamefont {Watanabe}}, \bibinfo {author} {\bibfnamefont
  {T.}~\bibnamefont {Taniguchi}}, \bibinfo {author} {\bibfnamefont
  {J.}~\bibnamefont {Huang}}, \bibinfo {author} {\bibfnamefont
  {M.}~\bibnamefont {Yi}}, \bibinfo {author} {\bibfnamefont {J.~S.}\
  \bibnamefont {Oh}},\ and\ \bibinfo {author} {\bibfnamefont {R.~J.}\
  \bibnamefont {Birgeneau}},\ }\href
  {https://doi.org/10.1021/acs.nanolett.1c04264} {\bibfield  {journal}
  {\bibinfo  {journal} {Nano Lett.}\ }\textbf {\bibinfo {volume} {22}},\
  \bibinfo {pages} {1151} (\bibinfo {year} {2022})}\BibitemShut {NoStop}%
\bibitem [{\citenamefont {Yang}\ \emph {et~al.}(2022)\citenamefont {Yang},
  \citenamefont {Liu}, \citenamefont {Zhou}, \citenamefont {Liu}, \citenamefont
  {Mu}, \citenamefont {Liu}, \citenamefont {Wang}, \citenamefont {Hao},
  \citenamefont {Li}, \citenamefont {Zhong}, \citenamefont {Du},\ and\
  \citenamefont {Zhuang}}]{Yang2022}%
  \BibitemOpen
  \bibfield  {author} {\bibinfo {author} {\bibfnamefont {M.}~\bibnamefont
  {Yang}}, \bibinfo {author} {\bibfnamefont {Y.}~\bibnamefont {Liu}}, \bibinfo
  {author} {\bibfnamefont {W.}~\bibnamefont {Zhou}}, \bibinfo {author}
  {\bibfnamefont {C.}~\bibnamefont {Liu}}, \bibinfo {author} {\bibfnamefont
  {D.}~\bibnamefont {Mu}}, \bibinfo {author} {\bibfnamefont {Y.}~\bibnamefont
  {Liu}}, \bibinfo {author} {\bibfnamefont {J.}~\bibnamefont {Wang}}, \bibinfo
  {author} {\bibfnamefont {W.}~\bibnamefont {Hao}}, \bibinfo {author}
  {\bibfnamefont {J.}~\bibnamefont {Li}}, \bibinfo {author} {\bibfnamefont
  {J.}~\bibnamefont {Zhong}}, \bibinfo {author} {\bibfnamefont
  {Y.}~\bibnamefont {Du}},\ and\ \bibinfo {author} {\bibfnamefont
  {J.}~\bibnamefont {Zhuang}},\ }\href
  {https://doi.org/10.1021/acsnano.1c10539} {\bibfield  {journal} {\bibinfo
  {journal} {ACS Nano}\ }\textbf {\bibinfo {volume} {16}},\ \bibinfo {pages}
  {3036} (\bibinfo {year} {2022})}\BibitemShut {NoStop}%
\bibitem [{\citenamefont {Zhong}\ \emph {et~al.}(2022)\citenamefont {Zhong},
  \citenamefont {Yang}, \citenamefont {Ye}, \citenamefont {Liu}, \citenamefont
  {Wang}, \citenamefont {Wang}, \citenamefont {Hao}, \citenamefont {Zhuang},\
  and\ \citenamefont {Du}}]{Zhong2022}%
  \BibitemOpen
  \bibfield  {author} {\bibinfo {author} {\bibfnamefont {J.}~\bibnamefont
  {Zhong}}, \bibinfo {author} {\bibfnamefont {M.}~\bibnamefont {Yang}},
  \bibinfo {author} {\bibfnamefont {F.}~\bibnamefont {Ye}}, \bibinfo {author}
  {\bibfnamefont {C.}~\bibnamefont {Liu}}, \bibinfo {author} {\bibfnamefont
  {J.}~\bibnamefont {Wang}}, \bibinfo {author} {\bibfnamefont {J.}~\bibnamefont
  {Wang}}, \bibinfo {author} {\bibfnamefont {W.}~\bibnamefont {Hao}}, \bibinfo
  {author} {\bibfnamefont {J.}~\bibnamefont {Zhuang}},\ and\ \bibinfo {author}
  {\bibfnamefont {Y.}~\bibnamefont {Du}},\ }\href
  {https://doi.org/10.1103/PhysRevApplied.17.064017} {\bibfield  {journal}
  {\bibinfo  {journal} {Phys. Rev. Appl.}\ }\textbf {\bibinfo {volume} {17}},\
  \bibinfo {pages} {064017} (\bibinfo {year} {2022})}\BibitemShut {NoStop}%
\bibitem [{\citenamefont {Zhang}\ \emph {et~al.}(2022)\citenamefont {Zhang},
  \citenamefont {Xing}, \citenamefont {Li}, \citenamefont {Peng}, \citenamefont
  {Qiao}, \citenamefont {Liu}, \citenamefont {Xiong}, \citenamefont {Han},
  \citenamefont {Liu}, \citenamefont {Xiao},\ and\ \citenamefont
  {Yao}}]{Zhang2022}%
  \BibitemOpen
  \bibfield  {author} {\bibinfo {author} {\bibfnamefont {X.}~\bibnamefont
  {Zhang}}, \bibinfo {author} {\bibfnamefont {X.}~\bibnamefont {Xing}},
  \bibinfo {author} {\bibfnamefont {J.}~\bibnamefont {Li}}, \bibinfo {author}
  {\bibfnamefont {X.}~\bibnamefont {Peng}}, \bibinfo {author} {\bibfnamefont
  {L.}~\bibnamefont {Qiao}}, \bibinfo {author} {\bibfnamefont {Y.}~\bibnamefont
  {Liu}}, \bibinfo {author} {\bibfnamefont {X.}~\bibnamefont {Xiong}}, \bibinfo
  {author} {\bibfnamefont {J.}~\bibnamefont {Han}}, \bibinfo {author}
  {\bibfnamefont {W.}~\bibnamefont {Liu}}, \bibinfo {author} {\bibfnamefont
  {W.}~\bibnamefont {Xiao}},\ and\ \bibinfo {author} {\bibfnamefont
  {Y.}~\bibnamefont {Yao}},\ }\href {https://doi.org/10.1063/5.0083807}
  {\bibfield  {journal} {\bibinfo  {journal} {Appl. Phys. Lett.}\ }\textbf
  {\bibinfo {volume} {120}},\ \bibinfo {pages} {093103} (\bibinfo {year}
  {2022})}\BibitemShut {NoStop}%
\bibitem [{\citenamefont {von Schnering}\ \emph {et~al.}(1978)\citenamefont
  {von Schnering}, \citenamefont {von Benda},\ and\ \citenamefont
  {Kalveram}}]{VonSchnering1978}%
  \BibitemOpen
  \bibfield  {author} {\bibinfo {author} {\bibfnamefont {H.~G.}\ \bibnamefont
  {von Schnering}}, \bibinfo {author} {\bibfnamefont {H.}~\bibnamefont {von
  Benda}},\ and\ \bibinfo {author} {\bibfnamefont {C.}~\bibnamefont
  {Kalveram}},\ }\href {https://doi.org/10.1002/zaac.19784380104} {\bibfield
  {journal} {\bibinfo  {journal} {Z. Anorg. Allg. Chem.}\ }\textbf {\bibinfo
  {volume} {438}},\ \bibinfo {pages} {37} (\bibinfo {year} {1978})}\BibitemShut
  {NoStop}%
\bibitem [{\citenamefont {Dikarev}\ \emph {et~al.}(2001)\citenamefont
  {Dikarev}, \citenamefont {Popovkin},\ and\ \citenamefont
  {Shevelkov}}]{Popovkin2001}%
  \BibitemOpen
  \bibfield  {author} {\bibinfo {author} {\bibfnamefont {E.~V.}\ \bibnamefont
  {Dikarev}}, \bibinfo {author} {\bibfnamefont {B.~A.}\ \bibnamefont
  {Popovkin}},\ and\ \bibinfo {author} {\bibfnamefont {A.~V.}\ \bibnamefont
  {Shevelkov}},\ }\href {https://doi.org/10.1023/A:1015010907973} {\bibfield
  {journal} {\bibinfo  {journal} {Russ. Chem. Bull.}\ }\textbf {\bibinfo
  {volume} {50}},\ \bibinfo {pages} {2304} (\bibinfo {year}
  {2001})}\BibitemShut {NoStop}%
\bibitem [{\citenamefont {Shimojima}\ \emph {et~al.}(2015)\citenamefont
  {Shimojima}, \citenamefont {Okazaki},\ and\ \citenamefont
  {Shin}}]{Shimojima2015}%
  \BibitemOpen
  \bibfield  {author} {\bibinfo {author} {\bibfnamefont {T.}~\bibnamefont
  {Shimojima}}, \bibinfo {author} {\bibfnamefont {K.}~\bibnamefont {Okazaki}},\
  and\ \bibinfo {author} {\bibfnamefont {S.}~\bibnamefont {Shin}},\ }\href
  {https://doi.org/10.7566/JPSJ.84.072001} {\bibfield  {journal} {\bibinfo
  {journal} {J. Phys. Soc. Japan}\ }\textbf {\bibinfo {volume} {84}},\ \bibinfo
  {pages} {072001} (\bibinfo {year} {2015})}\BibitemShut {NoStop}%
\bibitem [{\citenamefont {Yaji}\ \emph {et~al.}(2016)\citenamefont {Yaji},
  \citenamefont {Harasawa}, \citenamefont {Kuroda}, \citenamefont {Toyohisa},
  \citenamefont {Nakayama}, \citenamefont {Ishida}, \citenamefont {Fukushima},
  \citenamefont {Watanabe}, \citenamefont {Chen}, \citenamefont {Komori},\ and\
  \citenamefont {Shin}}]{Yaji2016}%
  \BibitemOpen
  \bibfield  {author} {\bibinfo {author} {\bibfnamefont {K.}~\bibnamefont
  {Yaji}}, \bibinfo {author} {\bibfnamefont {A.}~\bibnamefont {Harasawa}},
  \bibinfo {author} {\bibfnamefont {K.}~\bibnamefont {Kuroda}}, \bibinfo
  {author} {\bibfnamefont {S.}~\bibnamefont {Toyohisa}}, \bibinfo {author}
  {\bibfnamefont {M.}~\bibnamefont {Nakayama}}, \bibinfo {author}
  {\bibfnamefont {Y.}~\bibnamefont {Ishida}}, \bibinfo {author} {\bibfnamefont
  {A.}~\bibnamefont {Fukushima}}, \bibinfo {author} {\bibfnamefont
  {S.}~\bibnamefont {Watanabe}}, \bibinfo {author} {\bibfnamefont
  {C.}~\bibnamefont {Chen}}, \bibinfo {author} {\bibfnamefont {F.}~\bibnamefont
  {Komori}},\ and\ \bibinfo {author} {\bibfnamefont {S.}~\bibnamefont {Shin}},\
  }\href {https://doi.org/10.1063/1.4948738} {\bibfield  {journal} {\bibinfo
  {journal} {Rev. Sci. Instrum.}\ }\textbf {\bibinfo {volume} {87}},\ \bibinfo
  {pages} {053111} (\bibinfo {year} {2016})}\BibitemShut {NoStop}%
\bibitem [{\citenamefont {Dudin}\ \emph {et~al.}(2010)\citenamefont {Dudin},
  \citenamefont {Lacovig}, \citenamefont {Fava}, \citenamefont {Nicolini},
  \citenamefont {Bianco}, \citenamefont {Cautero},\ and\ \citenamefont
  {Barinov}}]{Dudin2010}%
  \BibitemOpen
  \bibfield  {author} {\bibinfo {author} {\bibfnamefont {P.}~\bibnamefont
  {Dudin}}, \bibinfo {author} {\bibfnamefont {P.}~\bibnamefont {Lacovig}},
  \bibinfo {author} {\bibfnamefont {C.}~\bibnamefont {Fava}}, \bibinfo {author}
  {\bibfnamefont {E.}~\bibnamefont {Nicolini}}, \bibinfo {author}
  {\bibfnamefont {A.}~\bibnamefont {Bianco}}, \bibinfo {author} {\bibfnamefont
  {G.}~\bibnamefont {Cautero}},\ and\ \bibinfo {author} {\bibfnamefont
  {A.}~\bibnamefont {Barinov}},\ }\href
  {https://doi.org/10.1107/S0909049510013993} {\bibfield  {journal} {\bibinfo
  {journal} {J. Synchrotron Radiat.}\ }\textbf {\bibinfo {volume} {17}},\
  \bibinfo {pages} {445} (\bibinfo {year} {2010})}\BibitemShut {NoStop}%
\bibitem [{\citenamefont {Blaha}\ \emph {et~al.}(2019)\citenamefont {Blaha},
  \citenamefont {Schwarz}, \citenamefont {Madsen}, \citenamefont {Kvasnicka},
  \citenamefont {Luitz}, \citenamefont {Laskowsk}, \citenamefont {Tran},
  \citenamefont {Marks},\ and\ \citenamefont {Marks}}]{Blaha2019}%
  \BibitemOpen
  \bibfield  {author} {\bibinfo {author} {\bibfnamefont {P.}~\bibnamefont
  {Blaha}}, \bibinfo {author} {\bibfnamefont {K.}~\bibnamefont {Schwarz}},
  \bibinfo {author} {\bibfnamefont {G.~K.~H.}\ \bibnamefont {Madsen}}, \bibinfo
  {author} {\bibfnamefont {D.}~\bibnamefont {Kvasnicka}}, \bibinfo {author}
  {\bibfnamefont {J.}~\bibnamefont {Luitz}}, \bibinfo {author} {\bibfnamefont
  {R.}~\bibnamefont {Laskowsk}}, \bibinfo {author} {\bibfnamefont
  {F.}~\bibnamefont {Tran}}, \bibinfo {author} {\bibfnamefont {L.}~\bibnamefont
  {Marks}},\ and\ \bibinfo {author} {\bibfnamefont {L.}~\bibnamefont {Marks}},\
  }\href@noop {} {\bibinfo {title} {{WIEN2k}}} (\bibinfo {year}
  {2019})\BibitemShut {NoStop}%
\bibitem [{sup()}]{supple}%
  \BibitemOpen
  \href@noop {} {\bibinfo  {journal} {See Supplemental Material at [URL] for
  the details of the crystal growth, structure analysis, and ARPES settings.}\
  }\BibitemShut {NoStop}%
\bibitem [{\citenamefont {Tang}\ \emph {et~al.}(2019)\citenamefont {Tang},
  \citenamefont {Po}, \citenamefont {Vishwanath},\ and\ \citenamefont
  {Wan}}]{Tang2019a}%
  \BibitemOpen
\bibfield  {journal} {  }\bibfield  {author} {\bibinfo {author} {\bibfnamefont
  {F.}~\bibnamefont {Tang}}, \bibinfo {author} {\bibfnamefont {H.~C.}\
  \bibnamefont {Po}}, \bibinfo {author} {\bibfnamefont {A.}~\bibnamefont
  {Vishwanath}},\ and\ \bibinfo {author} {\bibfnamefont {X.}~\bibnamefont
  {Wan}},\ }\href {https://doi.org/10.1038/s41586-019-0937-5} {\bibfield
  {journal} {\bibinfo  {journal} {Nature}\ }\textbf {\bibinfo {volume} {566}},\
  \bibinfo {pages} {486} (\bibinfo {year} {2019})}\BibitemShut {NoStop}%
\bibitem [{\citenamefont {Vergniory}\ \emph {et~al.}(2019)\citenamefont
  {Vergniory}, \citenamefont {Elcoro}, \citenamefont {Felser}, \citenamefont
  {Regnault}, \citenamefont {Bernevig},\ and\ \citenamefont
  {Wang}}]{Vergniory}%
  \BibitemOpen
  \bibfield  {author} {\bibinfo {author} {\bibfnamefont {M.~G.}\ \bibnamefont
  {Vergniory}}, \bibinfo {author} {\bibfnamefont {L.}~\bibnamefont {Elcoro}},
  \bibinfo {author} {\bibfnamefont {C.}~\bibnamefont {Felser}}, \bibinfo
  {author} {\bibfnamefont {N.}~\bibnamefont {Regnault}}, \bibinfo {author}
  {\bibfnamefont {B.~A.}\ \bibnamefont {Bernevig}},\ and\ \bibinfo {author}
  {\bibfnamefont {Z.}~\bibnamefont {Wang}},\ }\href
  {https://doi.org/10.1038/s41586-019-0954-4} {\bibfield  {journal} {\bibinfo
  {journal} {Nature}\ }\textbf {\bibinfo {volume} {566}},\ \bibinfo {pages}
  {480} (\bibinfo {year} {2019})}\BibitemShut {NoStop}%
\bibitem [{\citenamefont {Zhang}\ \emph {et~al.}(2019)\citenamefont {Zhang},
  \citenamefont {Jiang}, \citenamefont {Song}, \citenamefont {Huang},
  \citenamefont {He}, \citenamefont {Fang}, \citenamefont {Weng},\ and\
  \citenamefont {Fang}}]{Zhang2019l}%
  \BibitemOpen
  \bibfield  {author} {\bibinfo {author} {\bibfnamefont {T.}~\bibnamefont
  {Zhang}}, \bibinfo {author} {\bibfnamefont {Y.}~\bibnamefont {Jiang}},
  \bibinfo {author} {\bibfnamefont {Z.}~\bibnamefont {Song}}, \bibinfo {author}
  {\bibfnamefont {H.}~\bibnamefont {Huang}}, \bibinfo {author} {\bibfnamefont
  {Y.}~\bibnamefont {He}}, \bibinfo {author} {\bibfnamefont {Z.}~\bibnamefont
  {Fang}}, \bibinfo {author} {\bibfnamefont {H.}~\bibnamefont {Weng}},\ and\
  \bibinfo {author} {\bibfnamefont {C.}~\bibnamefont {Fang}},\ }\href
  {https://doi.org/10.1038/s41586-019-0944-6} {\bibfield  {journal} {\bibinfo
  {journal} {Nature}\ }\textbf {\bibinfo {volume} {566}},\ \bibinfo {pages}
  {475} (\bibinfo {year} {2019})}\BibitemShut {NoStop}%
\bibitem [{\citenamefont {Lv}\ \emph {et~al.}(2019)\citenamefont {Lv},
  \citenamefont {Qian},\ and\ \citenamefont {Ding}}]{Lv2019a}%
  \BibitemOpen
  \bibfield  {author} {\bibinfo {author} {\bibfnamefont {B.}~\bibnamefont
  {Lv}}, \bibinfo {author} {\bibfnamefont {T.}~\bibnamefont {Qian}},\ and\
  \bibinfo {author} {\bibfnamefont {H.}~\bibnamefont {Ding}},\ }\href
  {https://doi.org/10.1038/s42254-019-0088-5} {\bibfield  {journal} {\bibinfo
  {journal} {Nat. Rev. Phys.}\ }\textbf {\bibinfo {volume} {1}},\ \bibinfo
  {pages} {609} (\bibinfo {year} {2019})}\BibitemShut {NoStop}%
\end{thebibliography}
\end{document}